\documentstyle[11pt,epsf]{article}
\newfont{\subsub}{cmr6}

\newcounter{szk}

\begin{document}

\title{\bf The log-normal distribution from Non-Gibrat's law
in the middle scale region of profits 
%under detailed quasi-balance
}
\author{
\footnote{e-mail address: ishikawa@kanazawa-gu.ac.jp} 
Atushi Ishikawa
%$^a$ 
\\
%$^a$ 
Kanazawa Gakuin University, Kanazawa 920-1392, Japan
}
%\date{\today}
\date{}
\maketitle

%%%%%%%%%%%%%%%%%%%%%%%%%%%%%%%%%%%%%%%%%%%%%%
%%%       ABSTRACT
%%%%%%%%%%%%%%%%%%%%%%%%%%%%%%%%%%%%%%%%%%%%%%
\begin{abstract}
{
Employing profits data of Japanese firms in 2003--2005,
we kinematically exhibit the static log-normal distribution in the middle scale region.
In the derivation, a Non-Gibrat's law under the detailed balance is adopted
together with following two approximations.
Firstly, the probability density function of profits growth rate is described as 
a tent-shaped exponential function.
Secondly, the value of the origin of the growth rate distribution
divided into bins is constant. 
The derivation is confirmed in the database consistently.

This static procedure is applied to a quasi-static system.
We dynamically describe a quasi-static log-normal distribution in the middle scale region.
In the derivation, a Non-Gibrat's law under the detailed quasi-balance
is adopted together with two approximations confirmed in the static system.
The resultant distribution is power-law with varying Pareto index 
in the large scale region
and the quasi-static log-normal distribution in the middle scale region.
In the distribution,
not only the change of Pareto index 
but also the change of the variance of the log-normal distribution
depends on the parameter of the detailed quasi-balance.
As a result, Pareto index and 
the variance of the log-normal distribution are related to each other.

}
\end{abstract}
\begin{flushleft}
%Keywords : Econophysics; Profits; Log-normal distribution; 
%Non-Gibrat's law; Detailed quasi-balance
\end{flushleft}

%\vspace{1cm}
%%%%%%%%%%%%%%%%%%%%%%%%%%%%%%%%%%%%%%%%%%%%%%
%%%       SECTION
%%%       Introduction
%%%%%%%%%%%%%%%%%%%%%%%%%%%%%%%%%%%%%%%%%%%%%%\% 
\section{Introduction}
\label{sec-Introduction}
\indent

Log-normal distributions are frequently observed 
not only in natural phenomena but also in social phenomena.
For representative example, 
the probability density function $P(x)$ of personal income or
firm size $x$
is considered to obey the log-normal distribution~\cite{Gibrat}--\cite{Badger}
\begin{eqnarray}
    P_{\rm LN}(x) = \frac{1}{x \sqrt{2 \pi \sigma^2}} \exp 
        \left[-\frac{\ln^2 \left( x/\bar{x} \right)}{2 \sigma^2} \right]
    \label{LN}        
\end{eqnarray}
in the middle scale region.
\footnote{
In the large scale region, 
personal income distributions and firm size those
follow power-law \cite{Pareto}. We discuss this in Section 2.
In the low scale region, several distributions are proposed
(see Refs.~\cite{Yakovenko}--\cite{AIST} for instance).
We do not discuss them in this study.
}
Here $\bar{x}$ is a mean value and $\sigma^2$ is a variance.

The simplest model which describes the log-normal distribution
is the pure multiplicative stochastic process defined by
\begin{eqnarray}
    x(t+1) = R(t)~x(t)~,    
    \label{MSP}
\end{eqnarray}
where $R(t)$ is a positive random variable. 
By applying this process iteratively, we obtain
\begin{eqnarray}
    x(t) = R(t-1)~R(t-2)~\cdots~R(0)~x(0)~.  
    \label{MSP2}
\end{eqnarray}
The logarithm of this equation is
\begin{eqnarray}
    \log x(t) = \log R(t-1)+\log R(t-2)+\cdots+\log R(0)+\log x(0)~.  
    \label{MSP3}
\end{eqnarray}
If $\log x(0)$ is negligible compare to $\log x(t)$ in the limit $t \to \infty$ and

$\log R(i)$ $(i = 0, 1, \cdots, t-1)$ are independent probability variables, 
$\log x(t)$ follows the normal distribution in the limit.
As a result, this model derives the stationary log-normal distribution $P_{\rm LN} (x)$.

Eq.~(\ref{MSP}) means that the distribution of the growth rate
does not depend on $x(t)$.
This is known as Gibrat's law \cite{Gibrat}
that the conditional probability density function of the growth rate $Q(R|x_1)$
is independent of the initial value $x_1$:
\begin{eqnarray}
    Q(R|x_1) = Q(R)~.
    \label{Gibrat}
\end{eqnarray}
Here $x_1$ and $x_2$ are two successive incomes, assets, sales, profits,
the numbers of employees and so forth.
The growth rate $R$ is defined as $R=x_2/x_1$ and
$Q(R|x_1)$ as 
\begin{eqnarray}
    Q(R|x_1) = \frac{P_{1 R} (x_1, R) }{ P(x_1)}
    \label{define}
\end{eqnarray}
by using the probability density function $P(x_1)$ and the joint probability
density function $P_{1 R}(x_1, R)$.

As far as firm sizes in the middle scale region, however, it is reported that
the growth rate distributions do not follow the Gibrat's law (\ref{Gibrat}) 
(see Refs.~\cite{Stanley1}--\cite{Aoyama} for instance).
The pure multiplicative stochastic process model (\ref{MSP})
cannot be applied for explaining the log-normal distribution 
in the middle scale region.
Instead,
we have shown that the log-normal distribution can be derived
by using no model such as the pure multiplicative stochastic process~\cite{Ishikawa2007}.
In the derivation, two laws are employed which are observed 
in profits data of Japanese firms.
One is the law of detailed balance
which is observed in a stable economy \cite{FSAKA}.
The other is a Non-Gibrat's law 
which describes a statistical dependence 
in the growth rate of the past value~\cite{Ishikawa2007}.

In Ref.~\cite{Ishikawa2007},
we kinematically derived the static log-normal distribution under the detailed balance, and
analyzed empirical data in a single term 2003--2004
to confirm the derivation.
In this study, we show data analyses not only in 2003--2004 but also
in 2004--2005 and 2003--2005 to confirm the derivation more firmly.
After that, we dynamically derive the log-normal distribution
by replacing the detailed balance with the detailed quasi-balance proposed in Ref.~\cite{Ishikawa1}. 
By this procedure, the log-normal distribution is described as quasi-static.

%%%%%%%%%%%%%%%%%%%%%%%%%%%%%%%%%%%%%%%%%%%%%%%%%%
%%%       SECTION
%%%       Static log-normal distribution under the detailed balance
%%%%%%%%%%%%%%%%%%%%%%%%%%%%%%%%%%%%%%%%%%%%%%%%%%
\section{Static log-normal distribution under the detailed balance}
\label{Static log-normal distribution under the detailed balance}
\indent

In this section, we briefly review the study in Ref.~\cite{Ishikawa2007}.
After that, we empirically confirm the analytic result by added data analyses.
We employ profits data of Japanese firms in 2003, 2004 and 2005
which are available on the database
``CD Eyes 50" published in 2005 and 2006 by TOKYO SHOKO RESEARCH, LTD.~\cite{TSR}.

\begin{figure}[htb]
%%%%%%%%%%%%%%%%%%%%%%%%%%%%%%%%%%%%%%%
%%%%%%%% 1st Line (two pictures)%%%%%%%
%%%%%%%%%%%%%%%%%%%%%%%%%%%%%%%%%%%%%%%
 \begin{minipage}[htb]{0.49\textwidth}
  \epsfxsize = 1.0\textwidth
  \epsfbox{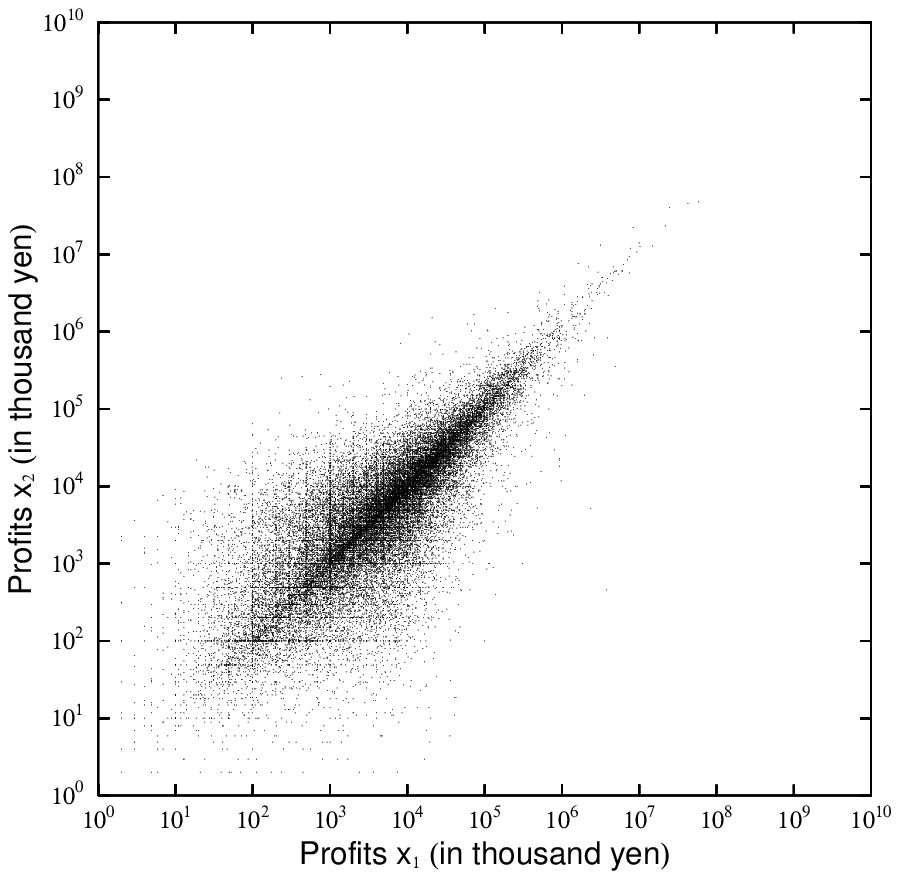}
 \caption{The scatter plot of all firms in the database,
 the profits of which in 2003 ($x_1$) and 2004 ($x_2$) exceeded $0$,
 $x_1 >0$ and $x_2 > 0$.
 The number of the firms is ``227,132".}
 \label{logProfit0304DB}
 \end{minipage}
 \hfill
 \begin{minipage}[htb]{0.49\textwidth}
  \epsfxsize = 1.0\textwidth
  \epsfbox{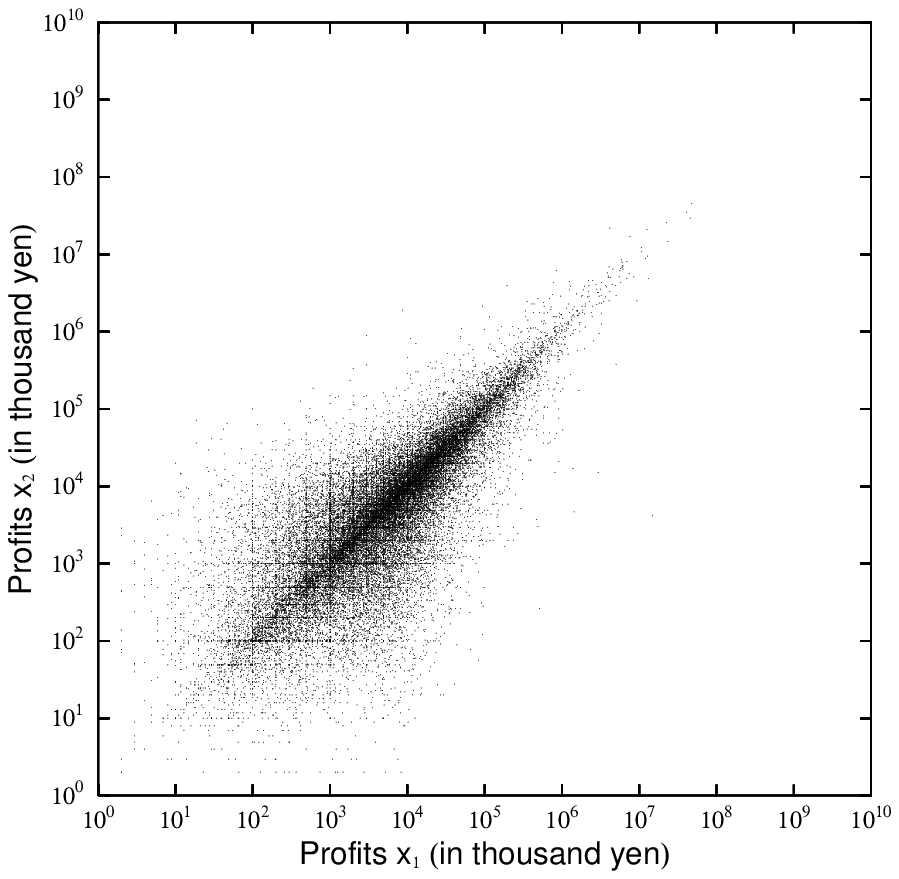}
 \caption{The scatter plot of all firms in the database,
 the profits of which in 2004 ($x_1$) and 2005 ($x_2$) exceeded $0$,
 $x_1 >0$ and $x_2 > 0$.
 The number of the firms is ``232,497 ".}
 \label{logProfit0405DB}
 \end{minipage}
\end{figure}

\begin{figure}[htb]
%%%%%%%%%%%%%%%%%%%%%%%%%%%%%%%%%%%%%%%
%%%%%%%% 1st Line (two pictures)%%%%%%%
%%%%%%%%%%%%%%%%%%%%%%%%%%%%%%%%%%%%%%%
%%%%%%%%%%%%%%%%%%%%%%%%%%%%%%%%%%%%%%%
 \begin{minipage}[htb]{0.49\textwidth}
  \epsfxsize = 1.0\textwidth
  \epsfbox{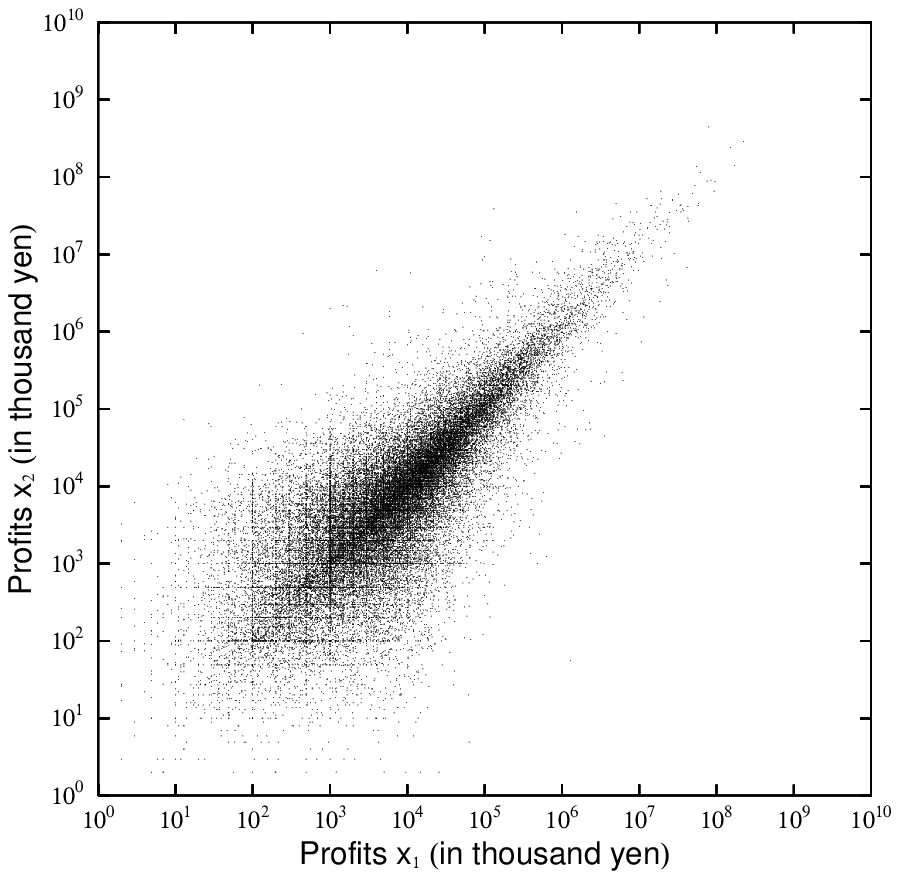}
 \caption{The scatter plot of all firms in the database,
 the profits of which in 2003 ($x_1$) and 2005 ($x_2$) exceeded $0$,
 $x_1 >0$ and $x_2 > 0$.
 The number of the firms is ``197,867 ".}
 \label{logProfit0305DB}
 \end{minipage}
 \hfill
 \begin{minipage}[htb]{0.49\textwidth}
  \epsfxsize = 1.0\textwidth
  \epsfbox{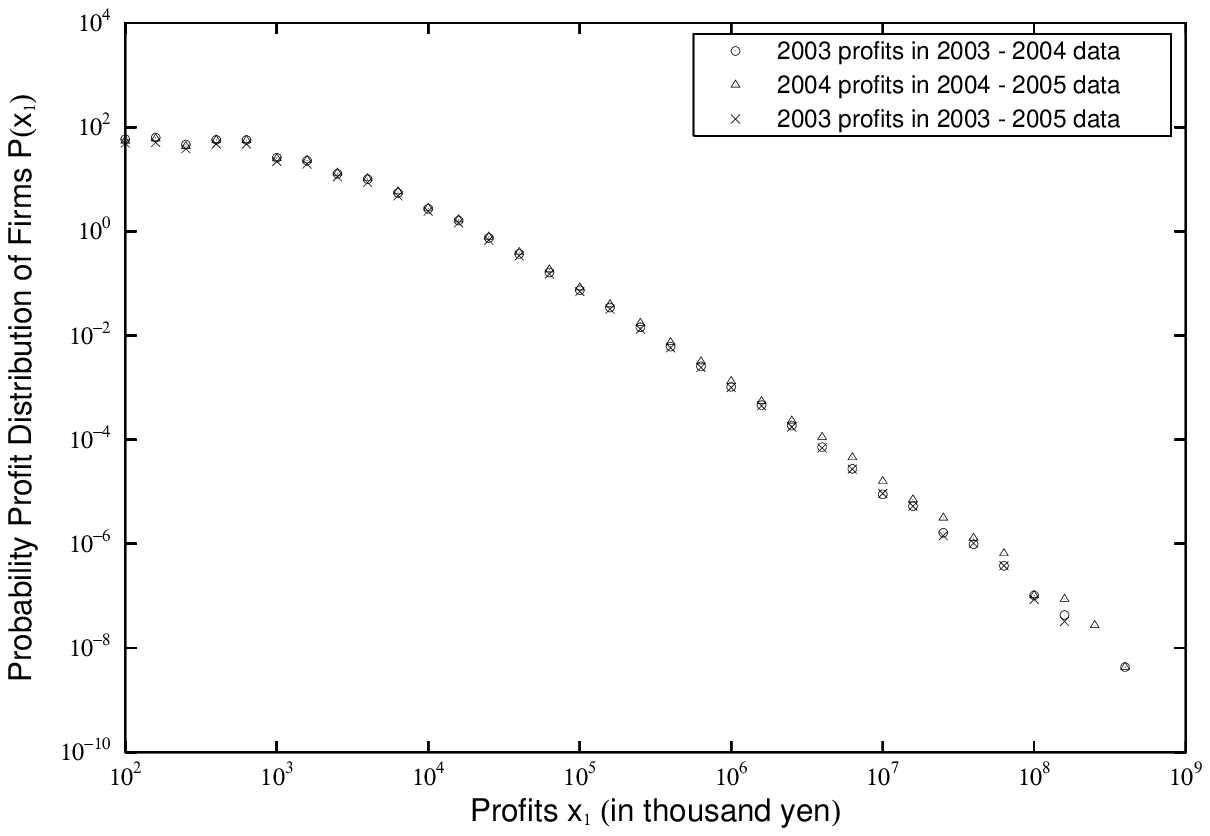}
 \caption{Probability density functions of profits 
in 2003--2005.
In the large scale region, Pareto's law is observed.
Pareto index is estimated to be nearly $1$.
}
 \label{DistributionData}
 \end{minipage}
\end{figure}

Figure \ref{logProfit0304DB} shows 
the joint probability density function $P_{1 2} (x_1, x_2)$
of all firms in the database,
the profits of which in 2003 ($x_1$)--2004 ($x_2$) exceeded $0$,
$x_1 >0$ and $x_2 > 0$.
The number of the firms is ``227,132".
Similarly, figures \ref{logProfit0405DB}--\ref{logProfit0305DB}
show the joint probability density functions
of the profits in 2004 ($x_1$)--2005 ($x_2$)
and 2003 ($x_1$)--2005 ($x_2$), 
and the number of the firms is ``232,497" and ``197,867", 
respectively. From Figs.~\ref{logProfit0304DB}--\ref{logProfit0305DB},
we approximately confirm the detailed balance
which is time-reversal symmetry ($x_1 \leftrightarrow x_2$) of 
$P_{1 2} (x_1, x_2)$ \cite{FSAKA}:
\begin{eqnarray}
    P_{1 2}(x_1, x_2) = P_{1 2}(x_2, x_1)~.
    \label{Detailed balance}
\end{eqnarray}

Figure \ref{DistributionData} shows probability density functions of profits 
in 2003--2005.
The distributions are almost stable and following power-law is observed
in the large scale region
\begin{eqnarray}
    P(x) = C x^{-\mu-1}~~~~{\rm for }~~~~x > x_0~,
    \label{Pareto}
\end{eqnarray}
where $x_0$ is a certain threshold.
This power-law is called Pareto's law \cite{Pareto}
and the exponent $\mu$ is named Pareto index.
%In this term, Pareto index is nearly $1$.
Notice that Pareto's law does not hold below the threshold $x_0$.
The purpose of this section is to exhibit the distribution in the middle scale region
under the detailed balance (\ref{Detailed balance}).

\begin{figure}[htb]
%%%%%%%%%%%%%%%%%%%%%%%%%%%%%%%%%%%%%%%
%%%%%%%% 1st Line (two pictures)%%%%%%%
%%%%%%%%%%%%%%%%%%%%%%%%%%%%%%%%%%%%%%%
 \begin{minipage}[htb]{0.49\textwidth}
  \epsfxsize = 1.0\textwidth
  \epsfbox{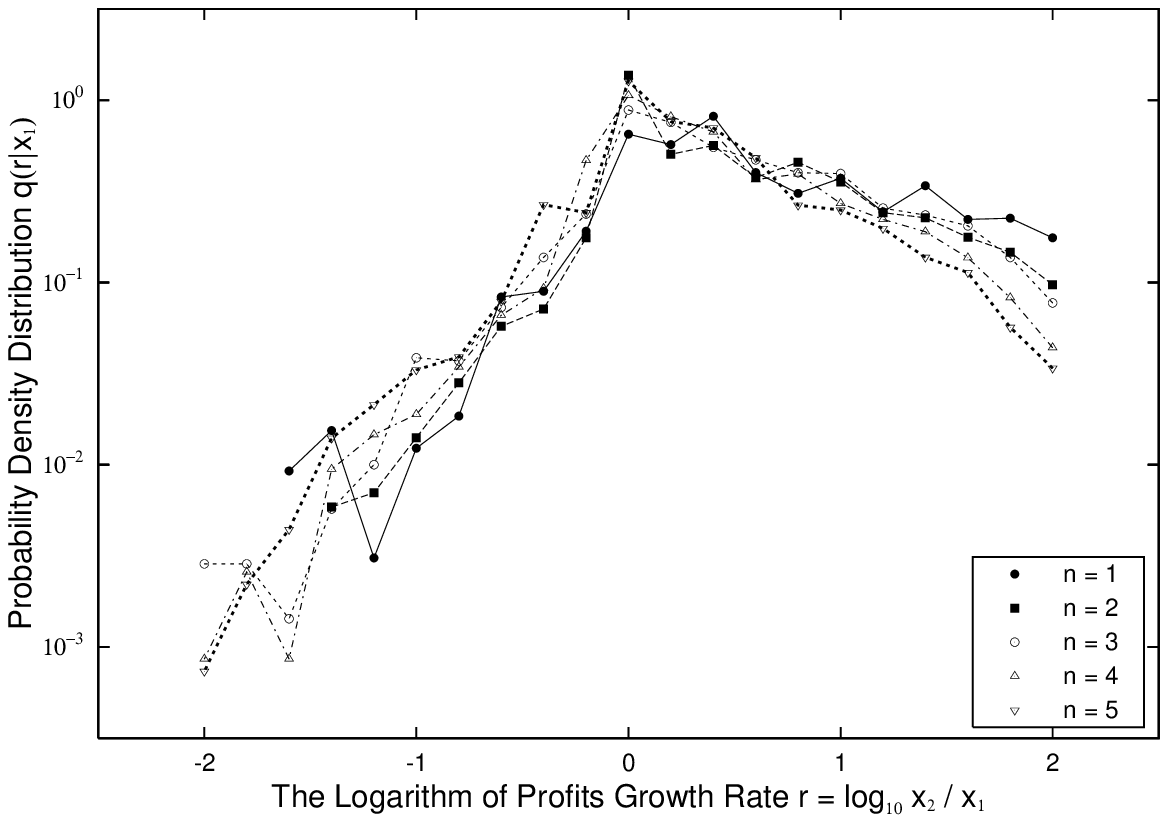}
 \end{minipage}
 \hfill
 \begin{minipage}[htb]{0.49\textwidth}
  \epsfxsize = 1.0\textwidth
  \epsfbox{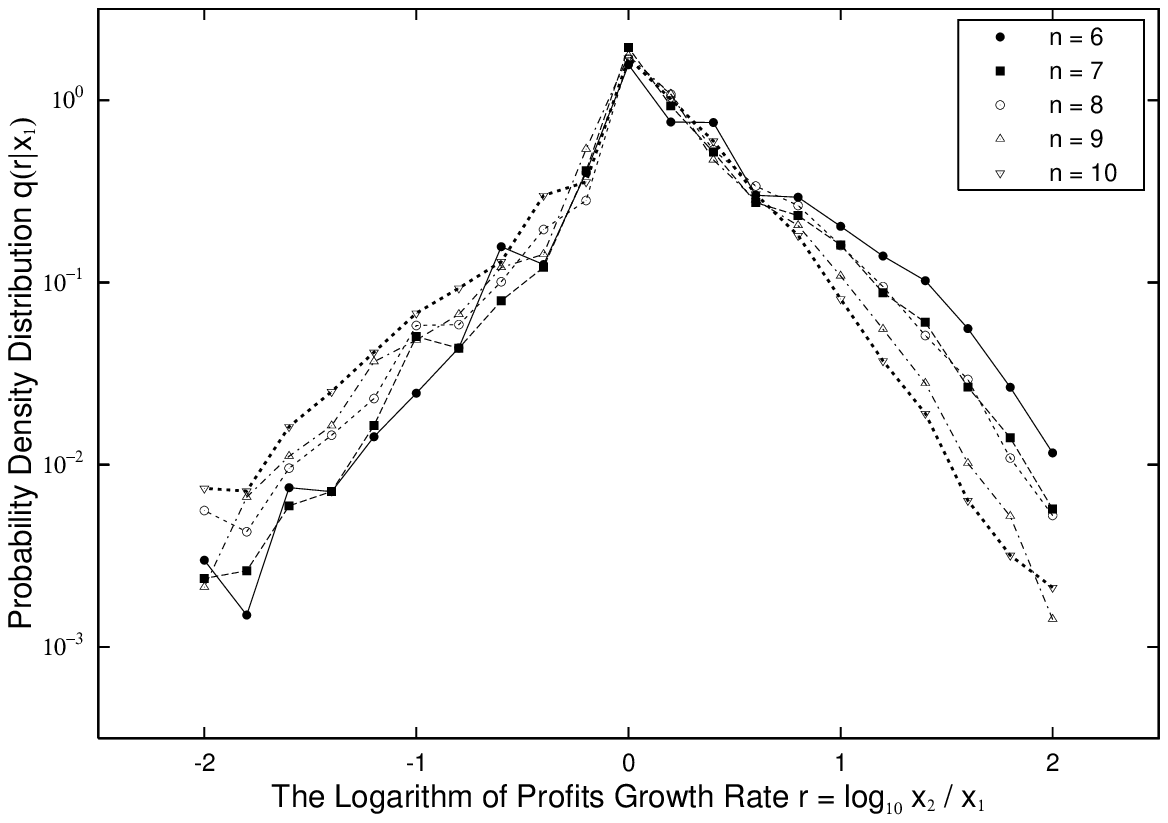}
 \end{minipage}
%\end{figure}
%%%%%%%%%%%% End 1st Line %%%%%%%%%%%%%
%%%%%%%%%%%%%%%%%%%%%%%%%%%%%%%%%%%%%%%
%%%%%%%% 2nd Line (two pictures)%%%%%%%
%%%%%%%%%%%%%%%%%%%%%%%%%%%%%%%%%%%%%%%
%\begin{figure}[htb]
 \begin{minipage}[htb]{0.49\textwidth}
  \epsfxsize = 1.0\textwidth
  \epsfbox{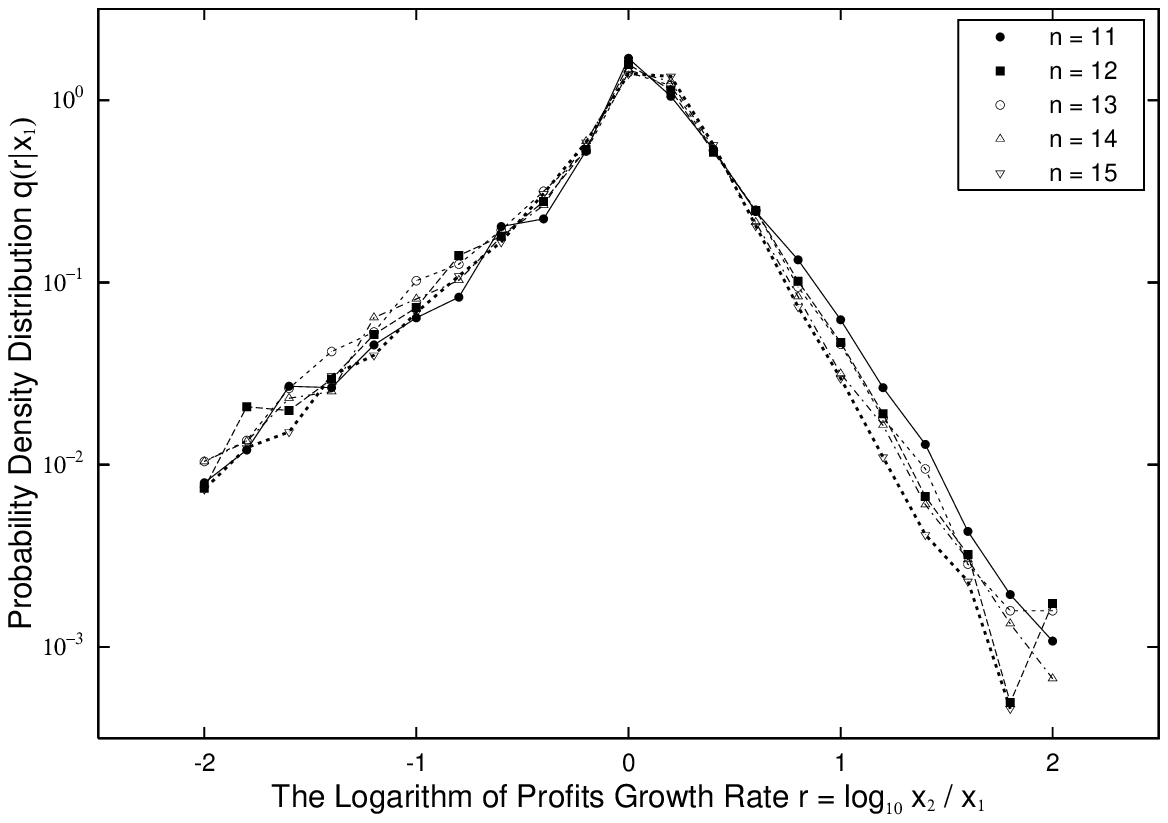}
 \end{minipage}
 \hfill
 \begin{minipage}[htb]{0.49\textwidth}
  \epsfxsize = 1.0\textwidth
  \epsfbox{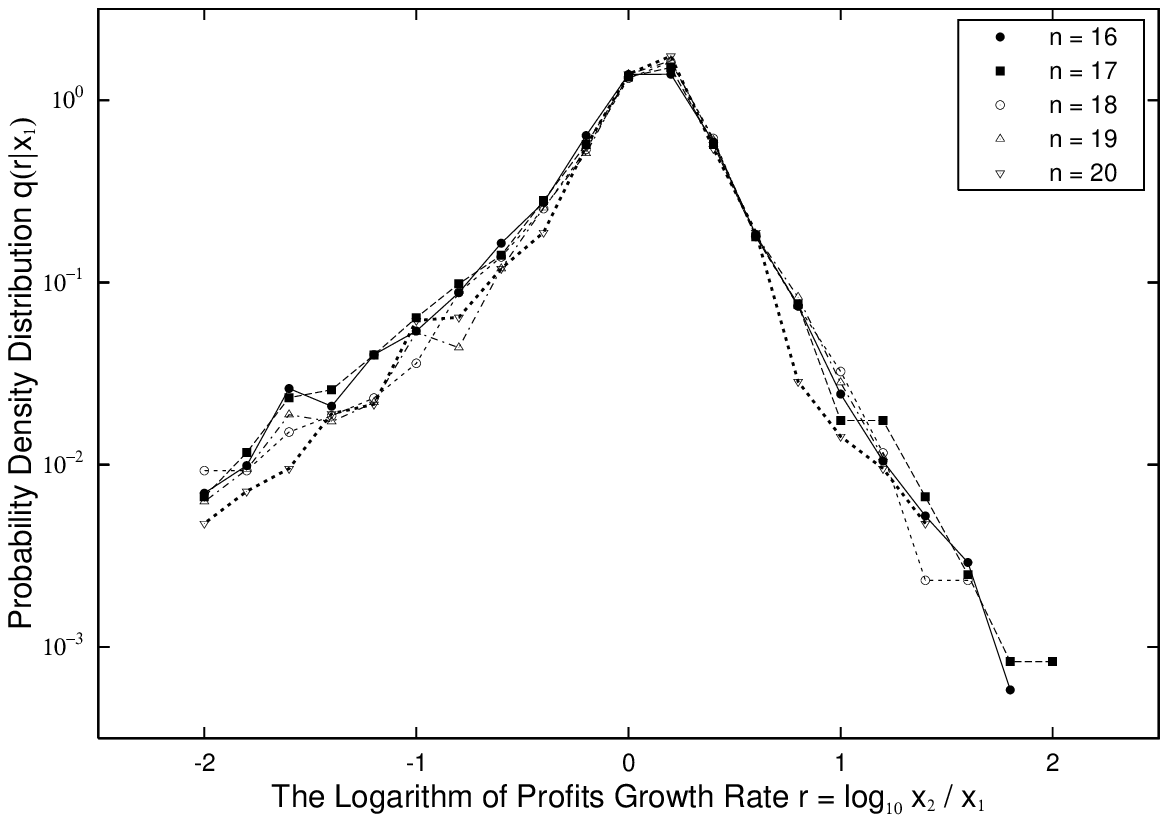}
 \end{minipage}
 \caption{The probability density distribution $q(r|x_1)$ of the log profits growth rate $r = \log_{10} x_2/x_1$ 
 from $2003$ to $2004$ for $n=1, 2, \cdots, 20$.}
 \label{ProfitGrowthRate0304}
\end{figure}

In order to identify a statistical dependence 
in the growth rate of the past value,
we examine the probability density faction of the profits growth rate in 2003--2004 firstly.
In the database, we divide the range of $x_1$ into logarithmically equal bins
as $x_1 \in 4 \times [10^{1+0.2(n-1)},10^{1+0.2n}]$ thousand yen with $n=1, 2, \cdots, 20$.
Figure \ref{ProfitGrowthRate0304} shows the probability density functions for $r = \log_{10} R$
in the case of $n=1, \cdots, 5$, $n=6, \cdots, 10$, $n=11, \cdots, 15$
and $n=16, \cdots, 20$.
The number of the firms in Fig.~\ref{ProfitGrowthRate0304}
is ``$22,005$", ``$89,507$", ``$85,020$" and ``$24,203$", respectively.

From Fig.~\ref{ProfitGrowthRate0304},
we approximate $\log_{10} q(r|x_1)$ by linear functions of $r$:
\begin{eqnarray}
    \log_{10}q(r|x_1)&=&c(x_1)-t_{+}(x_1)~r~~~~~{\rm for}~~r > 0~,
    \label{approximation1}\\
    \log_{10}q(r|x_1)&=&c(x_1)+t_{-}(x_1)~r~~~~~{\rm for}~~r < 0~.
    \label{approximation2}
\end{eqnarray}
Here $q(r|x_1)$ is the probability density function for $r$,
which is related to that for $R$ by $Q(R|x_1) = \log_{10}q(r|x_1) - r - \log_{10}(\ln 10)$.
These approximations (\ref{approximation1})--(\ref{approximation2})
are expressed as tent-shaped exponential forms as follows
\begin{eqnarray}
    Q(R|x_1)&=&d(x_1)~R^{-t_{+}(x_1)-1}~~~~~{\rm for}~~R > 1~,
    \label{tent-shaped1}\\
    Q(R|x_1)&=&d(x_1)~R^{+t_{-}(x_1)-1}~~~~~{\rm for}~~R < 1~,
    \label{tent-shaped2}
\end{eqnarray}
where $d(x_1)=10^{c(x_1)}/{\ln 10}$.
Furthermore, figure~\ref{ProfitGrowthRate0304} shows that
the dependence of $c(x_1)$ on $x_1$ is negligible for $n = 9, \cdots, 20$~.
We assess the validity of these approximations against the results.

\begin{figure}[htb]
%%%%%%%%%%%%%%%%%%%%%%%%%%%%%%%%%%%%%%%
%%%%%%%% 1st Line (two pictures)%%%%%%%
%%%%%%%%%%%%%%%%%%%%%%%%%%%%%%%%%%%%%%%
 \begin{minipage}[htb]{0.49\textwidth}
  \epsfxsize = 1.0\textwidth
  \epsfbox{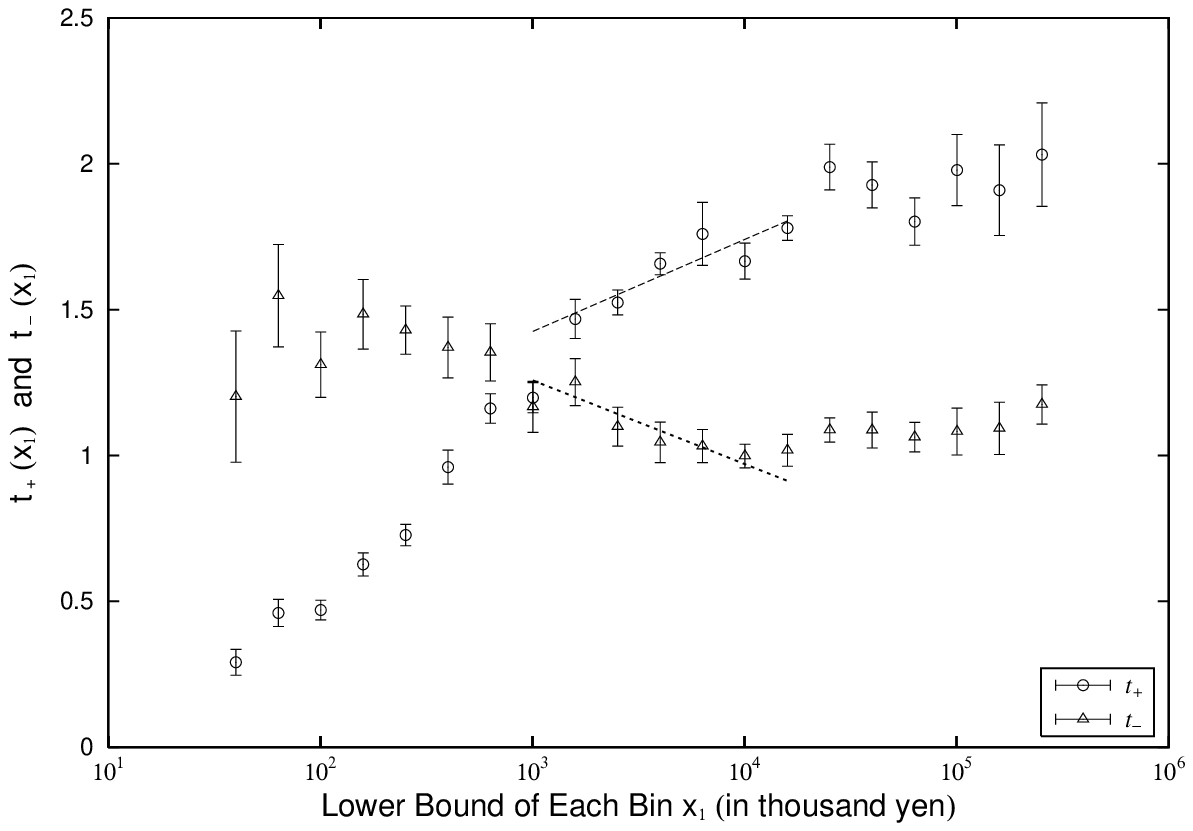}
 \caption{The relation between the lower bound of each bin $x_1$ and $t_{\pm}(x_1)$
 with respect to the profits growth rate from $2003$ to $2004$.
 From the left, each data point represents $n=1, 2, \cdots, 20$.}
 \label{eGibrat0304}
 \end{minipage}
 \hfill
 \begin{minipage}[htb]{0.49\textwidth}
  \epsfxsize = 1.0\textwidth
  \epsfbox{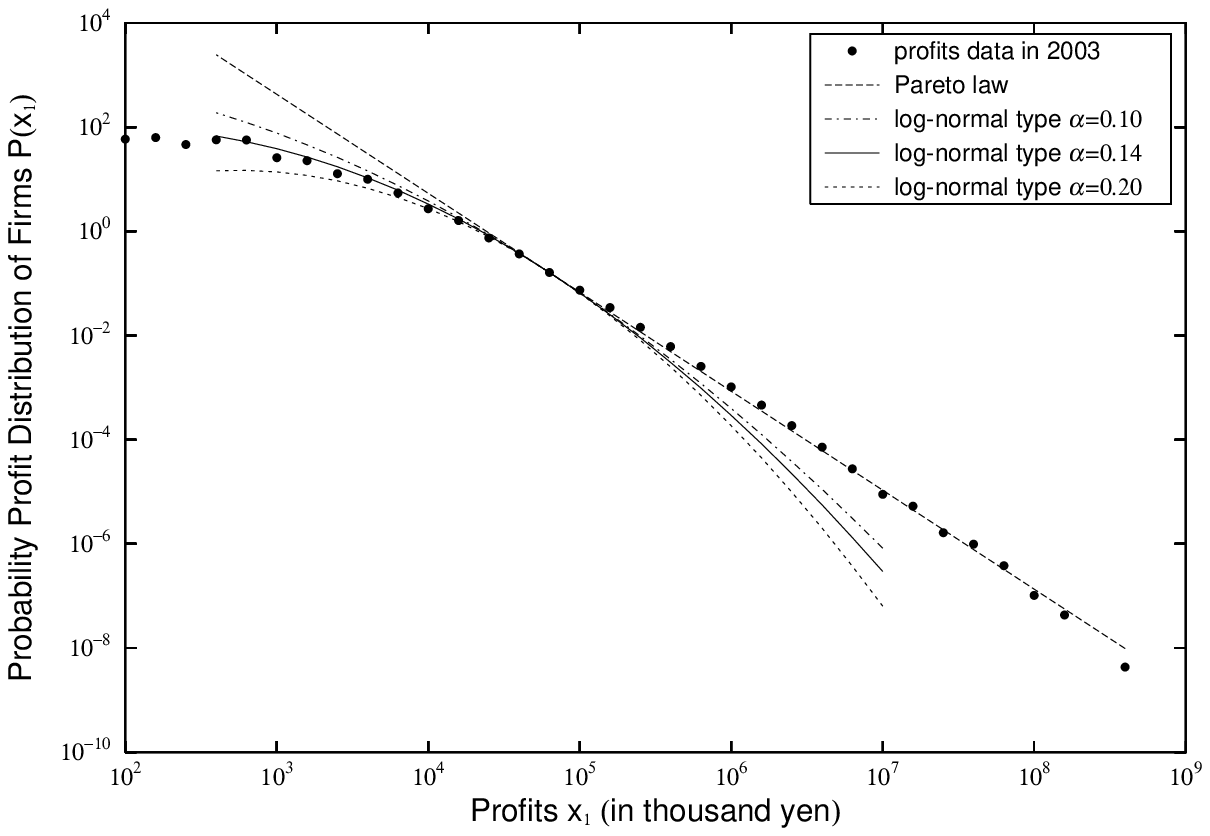}
 \caption{The probability distribution function of profits $P(x_1)$ for firms,
 the profits of which in 2003 ($x_1$) and 2004 ($x_2$) exceeded $0$,
 $x_1 >0$ and $x_2 > 0$.
 }
 \vspace{6.5mm}
 \label{Distribution0304-03}
 \end{minipage}
\end{figure}

Figure~\ref{eGibrat0304} represents
the dependence of $t_{\pm}(x_1)$ on the lower bound of each bin
$x_1 = 4 \times 10^{1+0.2(n-1)}$.
For $n=17, \cdots, 20$, $t_{\pm}(x_1)$ hardly responds to $x_1$.
This means that Gibrat's law (\ref{Gibrat}) holds only 
in the large scale region of profits ($x > x_0$). \footnote{
Fujiwara et al.~\cite{FSAKA} prove that Pareto's law (\ref{Pareto}) is derived
from the Gibrat's law (\ref{Gibrat}) valid only in the large scale region
and the detailed balance (\ref{Detailed balance}).
In the derivation, linear approximations (\ref{approximation1})--(\ref{approximation2})
need not be assumed.
}
In contrast, $t_{+}(x_1)$ linearly increases and $t_{-}(x_1)$ linearly decreases 
symmetrically with $\log_{10} x_1$ for $n = 9, \cdots, 13$~. From
Fig.~\ref{eGibrat0304}, the slops are described as~\cite{Ishikawa3} 
\begin{eqnarray}
    t_{\pm}(x_1)=t_{\pm}(x_0) \pm \alpha~\ln \frac{x_1}{x_0}~.
    \label{t0}
\end{eqnarray}
The parameters are estimated as follows: 
$\alpha \sim 0$ for $x_1 > x_0$, 
$\alpha \sim 0.14$ for $x_{\rm min} < x_1 < x_0$,
$x_0 = 4 \times 10^{1+0.2(17-1)} \sim 63,000$ thousand yen and
$x_{\rm min} = 4 \times 10^{1+0.2(9-1)} \sim 1,600$ thousand yen.
Notice that 
approximations (\ref{approximation1})--(\ref{approximation2})
uniquely fix
the expression of $t_{\pm}(x_1)$
under the detailed balance~\cite{Ishikawa2007}.
This derivation is included in the proof in the next section.
We call Eqs.(\ref{tent-shaped1})--(\ref{t0}) Non-Gibrat's law.

For $n = 9, \cdots, 20$, the dependence of $d(x_1)$ on $x_1$ is negligible.
In this case, the Non-Gibrat's law determines the probability density function of profits
as follows
\begin{eqnarray}
    P(x_1) = C {x_1}^{-\left(\mu+1\right)}~e^{-\alpha \ln^2 \frac{x_1}{x_0}}
    ~~~~~{\rm for}~~x_1 > x_{\rm min}~,
    \label{HandM}
\end{eqnarray}
where $t_{+}(x_0) - t_{-}(x_0) \sim \mu$~\cite{Ishikawa0}.
This is the power-law in the large scale region ($x_1 > x_0$)
and the log-normal distribution in the middle scale region 
($x_{\rm min} < x_1 < x_0$).
The relations between parameters $\sigma$, $\bar{x}$ in Eq.~(\ref{LN})
and $\alpha$, $\mu$, $x_0$ are given by
\begin{eqnarray}
    \alpha = \frac{1}{2 \sigma^2}~,~~~\mu = \frac{1}{\sigma^2} \ln \frac{x_0}{\bar{x}}~.
\end{eqnarray}
Figure~\ref{Distribution0304-03} shows that
the distribution (\ref{HandM}) fits with the empirical data consistently.
Notice that the distribution cannot fit with the empirical data,
if $\alpha$ is different from the value estimated in Fig.~\ref{eGibrat0304}
($\alpha = 0.10$ or $\alpha = 0.20$ for instance).

\begin{figure}[htb]
%%%%%%%%%%%%%%%%%%%%%%%%%%%%%%%%%%%%%%%
%%%%%%%% 1st Line (two pictures)%%%%%%%
%%%%%%%%%%%%%%%%%%%%%%%%%%%%%%%%%%%%%%%
 \begin{minipage}[htb]{0.49\textwidth}
  \epsfxsize = 1.0\textwidth
  \epsfbox{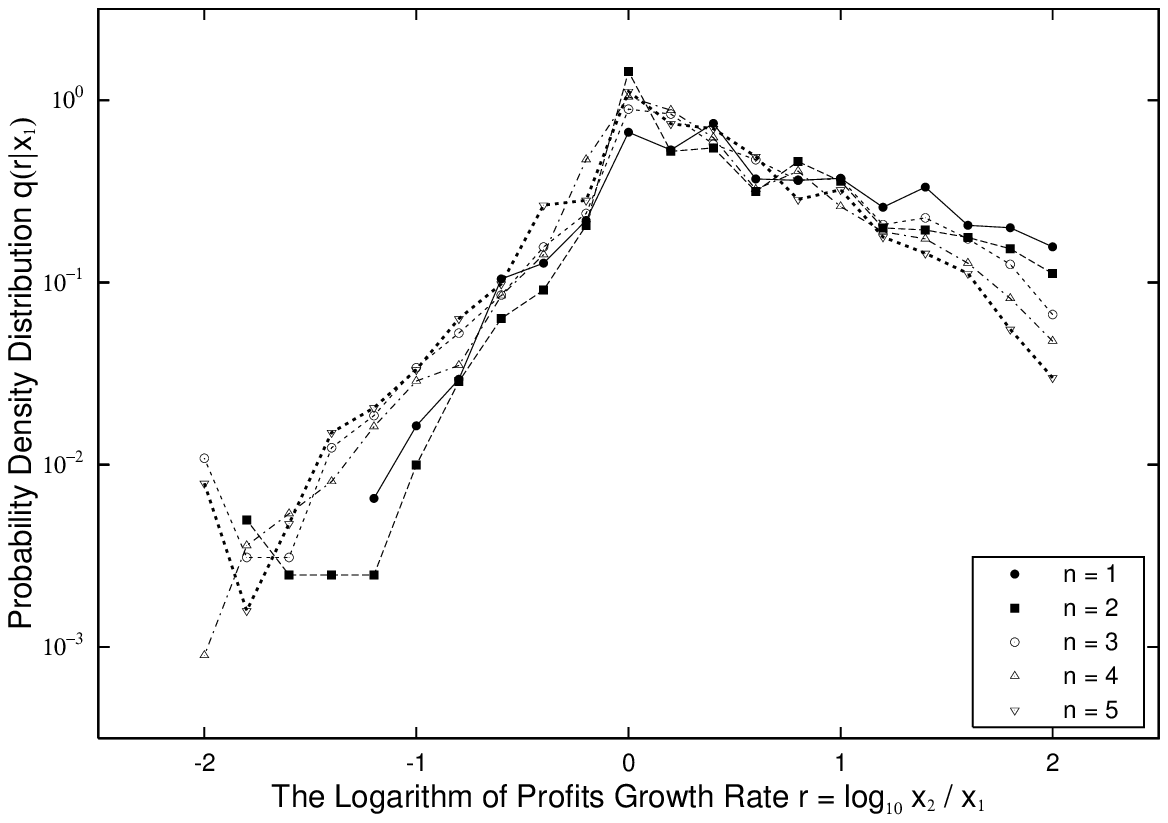}
 \end{minipage}
 \hfill
 \begin{minipage}[htb]{0.49\textwidth}
  \epsfxsize = 1.0\textwidth
  \epsfbox{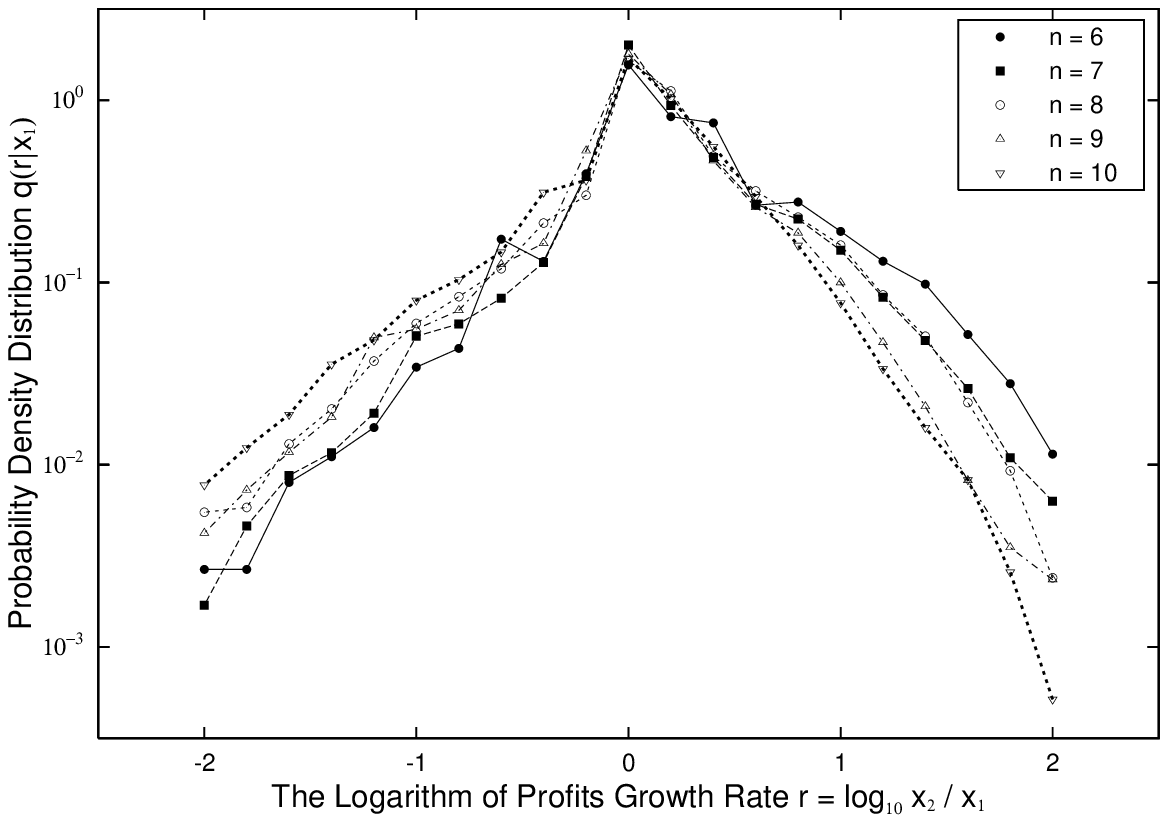}
 \end{minipage}
%\end{figure}
%%%%%%%%%%%% End 1st Line %%%%%%%%%%%%%
%%%%%%%%%%%%%%%%%%%%%%%%%%%%%%%%%%%%%%%
%%%%%%%% 2nd Line (two pictures)%%%%%%%
%%%%%%%%%%%%%%%%%%%%%%%%%%%%%%%%%%%%%%%
%\begin{figure}[htb]
 \begin{minipage}[htb]{0.49\textwidth}
  \epsfxsize = 1.0\textwidth
  \epsfbox{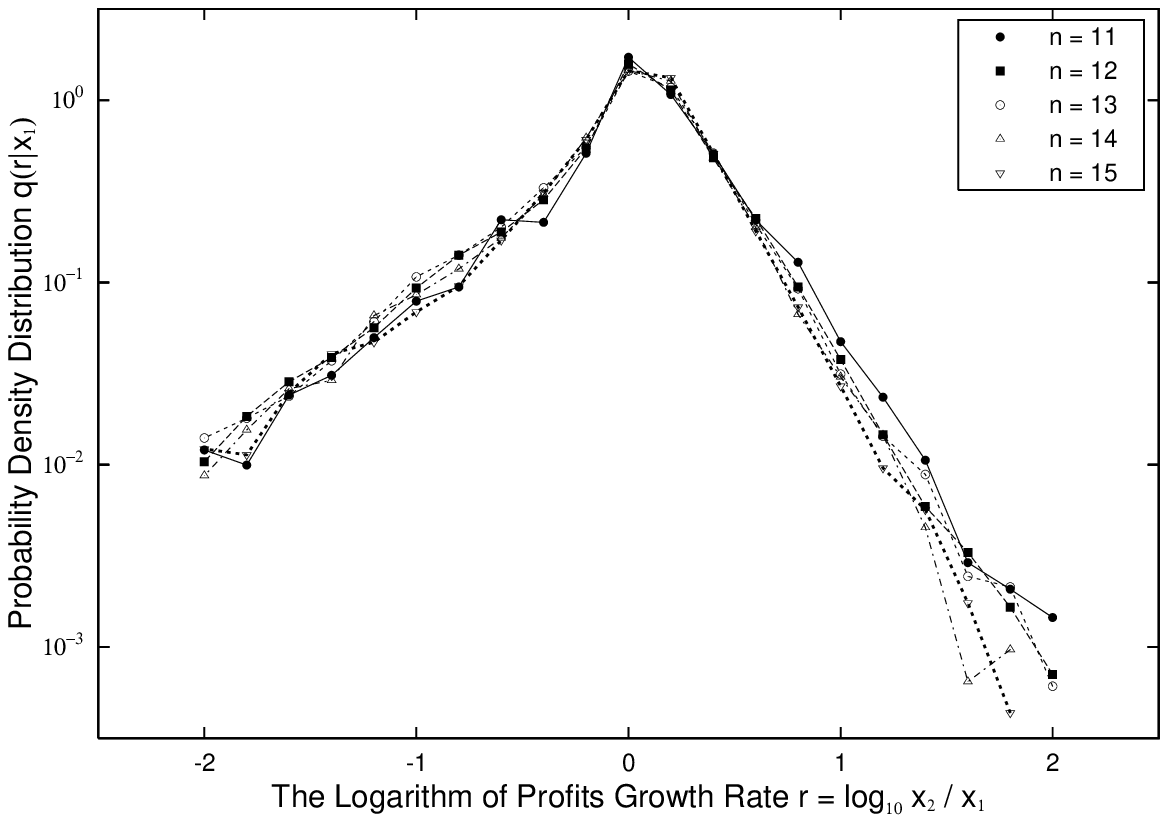}
 \end{minipage}
 \hfill
 \begin{minipage}[htb]{0.49\textwidth}
  \epsfxsize = 1.0\textwidth
  \epsfbox{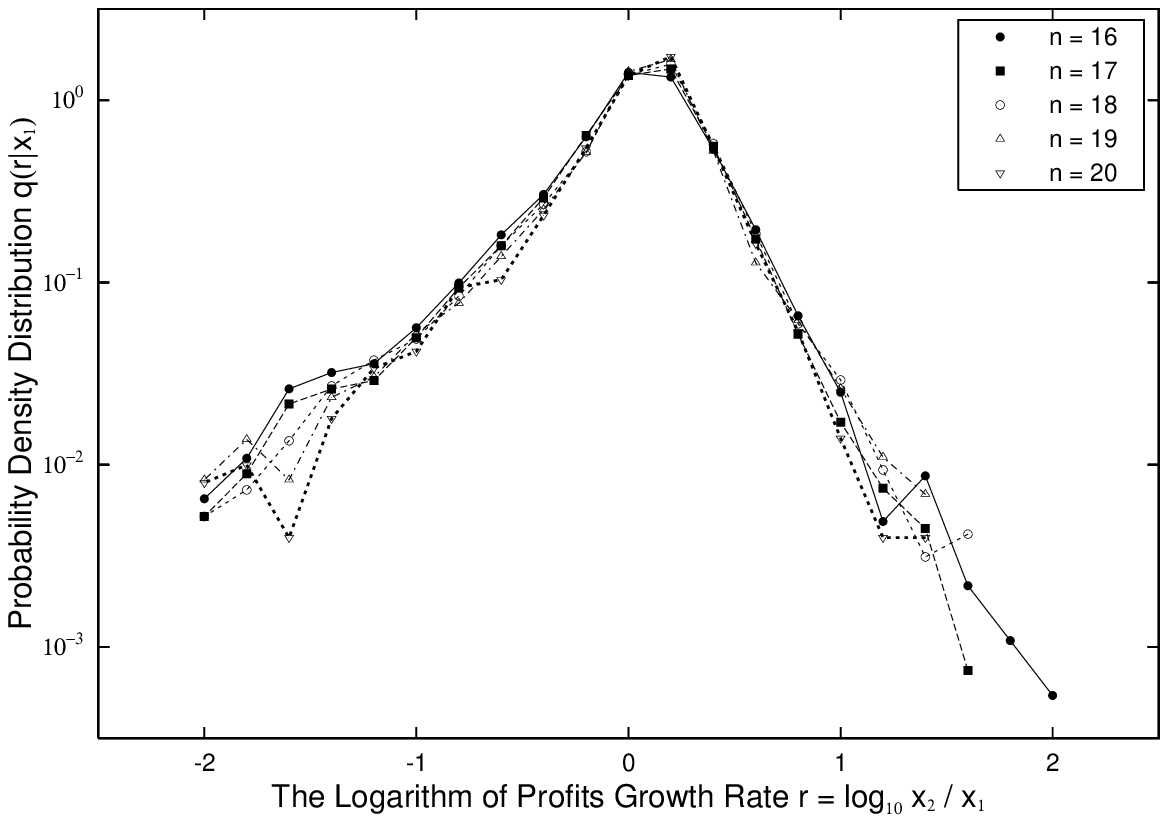}
 \end{minipage}
 \caption{The probability density distribution $q(r|x_1)$ of the log profits growth rate 
 $r = \log_{10} x_2/x_1$ 
 from $2004$ to $2005$ for $n=1, 2, \cdots, 20$.}
 \label{ProfitGrowthRate0405}
\end{figure}

\begin{figure}[htb]
%%%%%%%%%%%%%%%%%%%%%%%%%%%%%%%%%%%%%%%
%%%%%%%% 1st Line (two pictures)%%%%%%%
%%%%%%%%%%%%%%%%%%%%%%%%%%%%%%%%%%%%%%%
 \begin{minipage}[htb]{0.49\textwidth}
  \epsfxsize = 1.0\textwidth
  \epsfbox{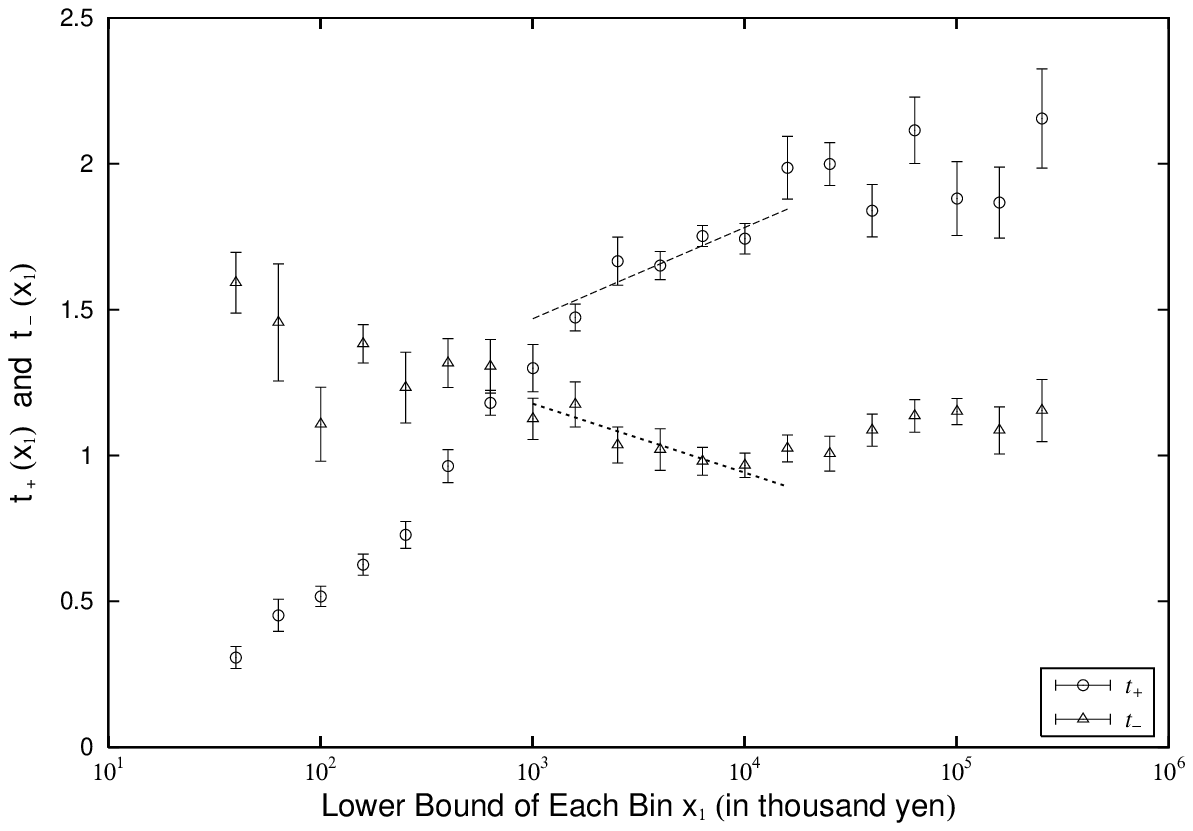}
 \caption{The relation between the lower bound of each bin $x_1$ and $t_{\pm}(x_1)$
 with respect to the profits growth rate from $2004$ to $2005$.
 From the left, each data point represents $n=1, 2, \cdots, 20$.}
 \label{eGibrat0405}
 \end{minipage}
 \hfill
 \begin{minipage}[htb]{0.49\textwidth}
  \epsfxsize = 1.0\textwidth
  \epsfbox{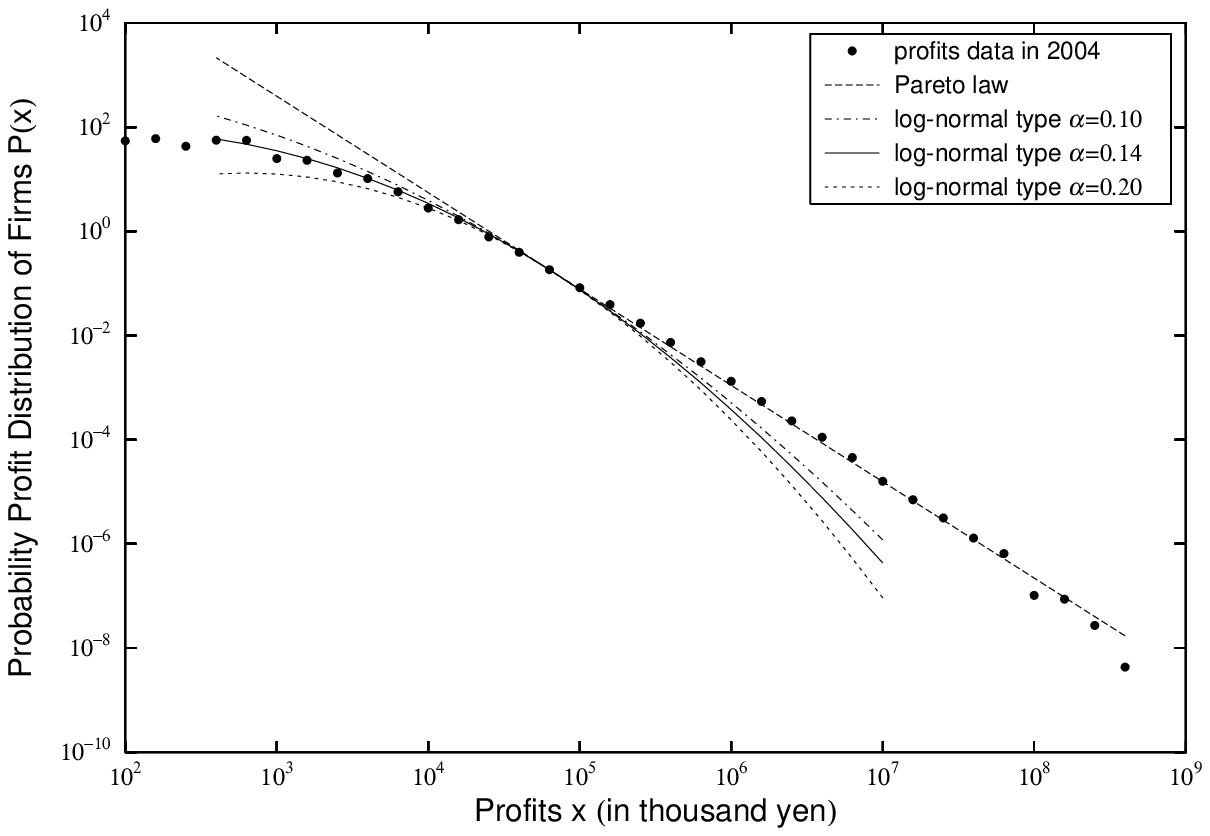}
 \caption{The probability distribution function of profits $P(x_1)$ for firms,
 the profits of which in 2004 ($x_1$) and 2005 ($x_2$) exceeded $0$,
 $x_1 >0$ and $x_2 > 0$.
 }
 \vspace{6.5mm}
 \label{Distribution0405-04}
 \end{minipage}
\end{figure}

\begin{figure}[htb]
%%%%%%%%%%%%%%%%%%%%%%%%%%%%%%%%%%%%%%%
%%%%%%%% 1st Line (two pictures)%%%%%%%
%%%%%%%%%%%%%%%%%%%%%%%%%%%%%%%%%%%%%%%
 \begin{minipage}[htb]{0.49\textwidth}
  \epsfxsize = 1.0\textwidth
  \epsfbox{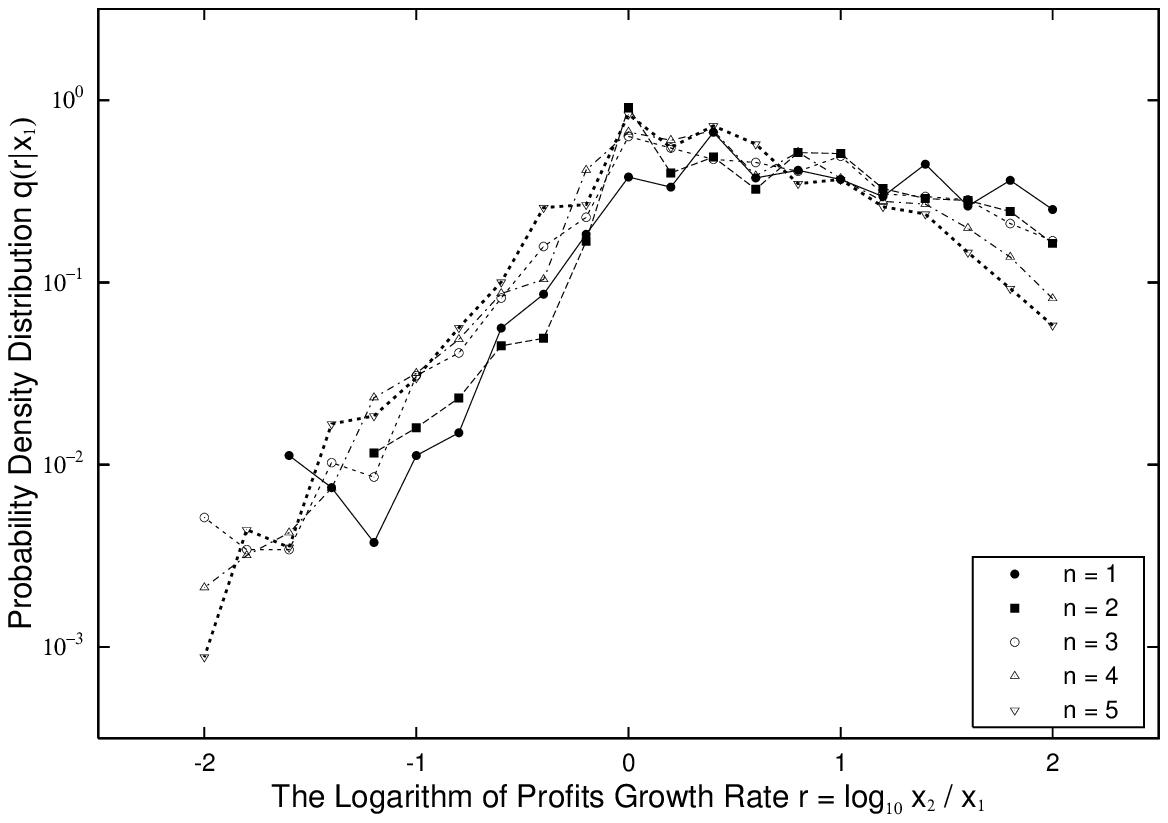}
 \end{minipage}
 \hfill
 \begin{minipage}[htb]{0.49\textwidth}
  \epsfxsize = 1.0\textwidth
  \epsfbox{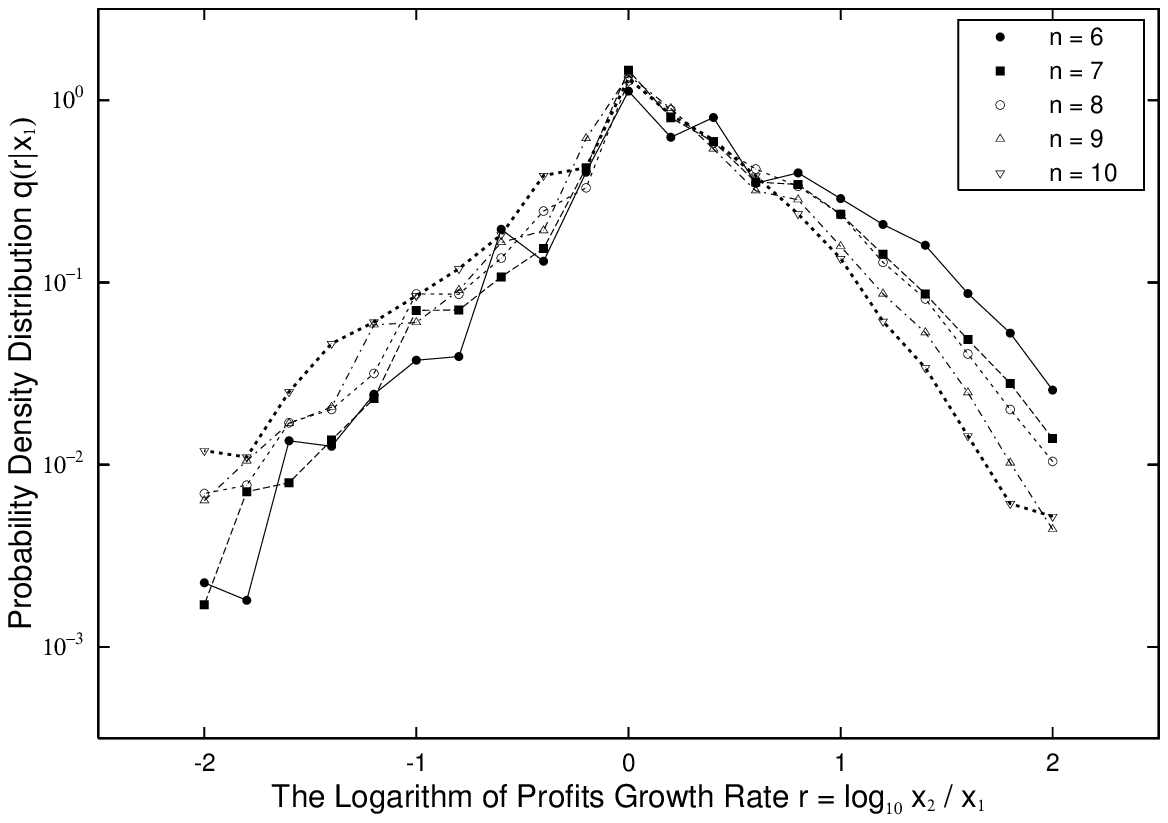}
 \end{minipage}
%\end{figure}
%%%%%%%%%%%% End 1st Line %%%%%%%%%%%%%
%%%%%%%%%%%%%%%%%%%%%%%%%%%%%%%%%%%%%%%
%%%%%%%% 2nd Line (two pictures)%%%%%%%
%%%%%%%%%%%%%%%%%%%%%%%%%%%%%%%%%%%%%%%
%\begin{figure}[htb]
 \begin{minipage}[htb]{0.49\textwidth}
  \epsfxsize = 1.0\textwidth
  \epsfbox{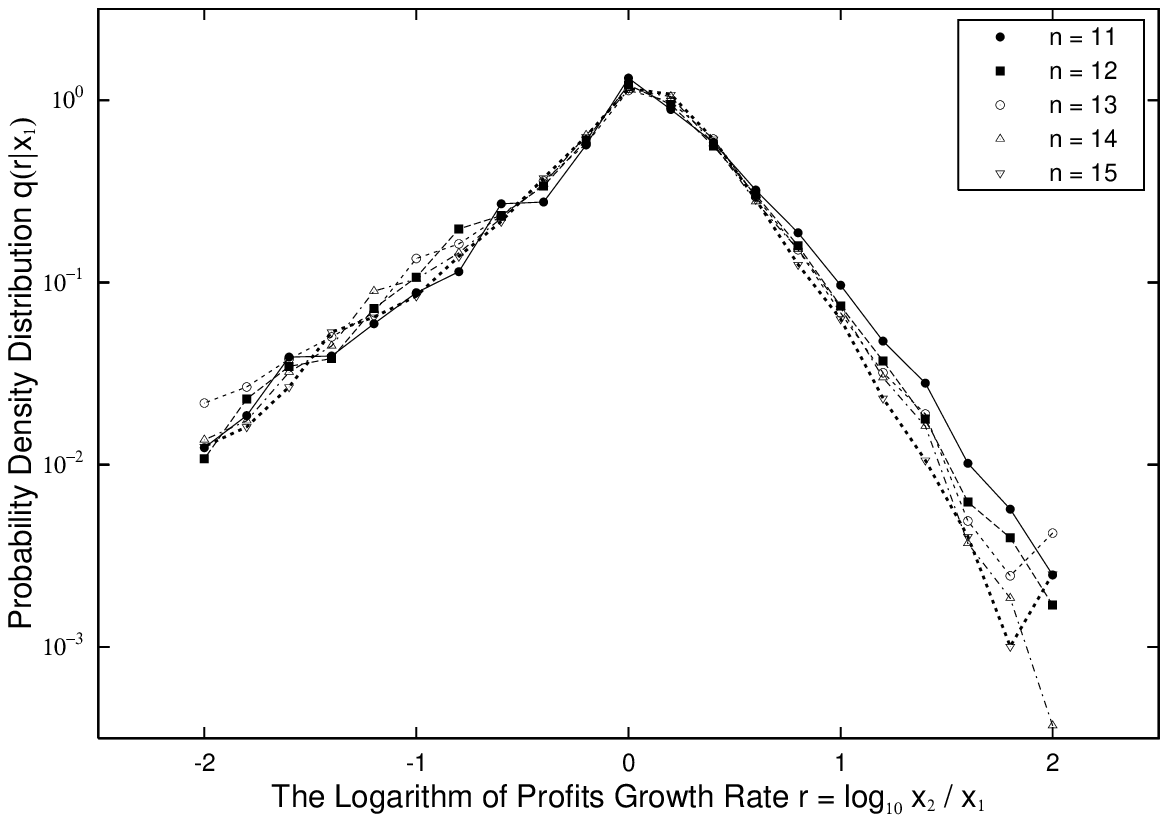}
 \end{minipage}
 \hfill
 \begin{minipage}[htb]{0.49\textwidth}
  \epsfxsize = 1.0\textwidth
  \epsfbox{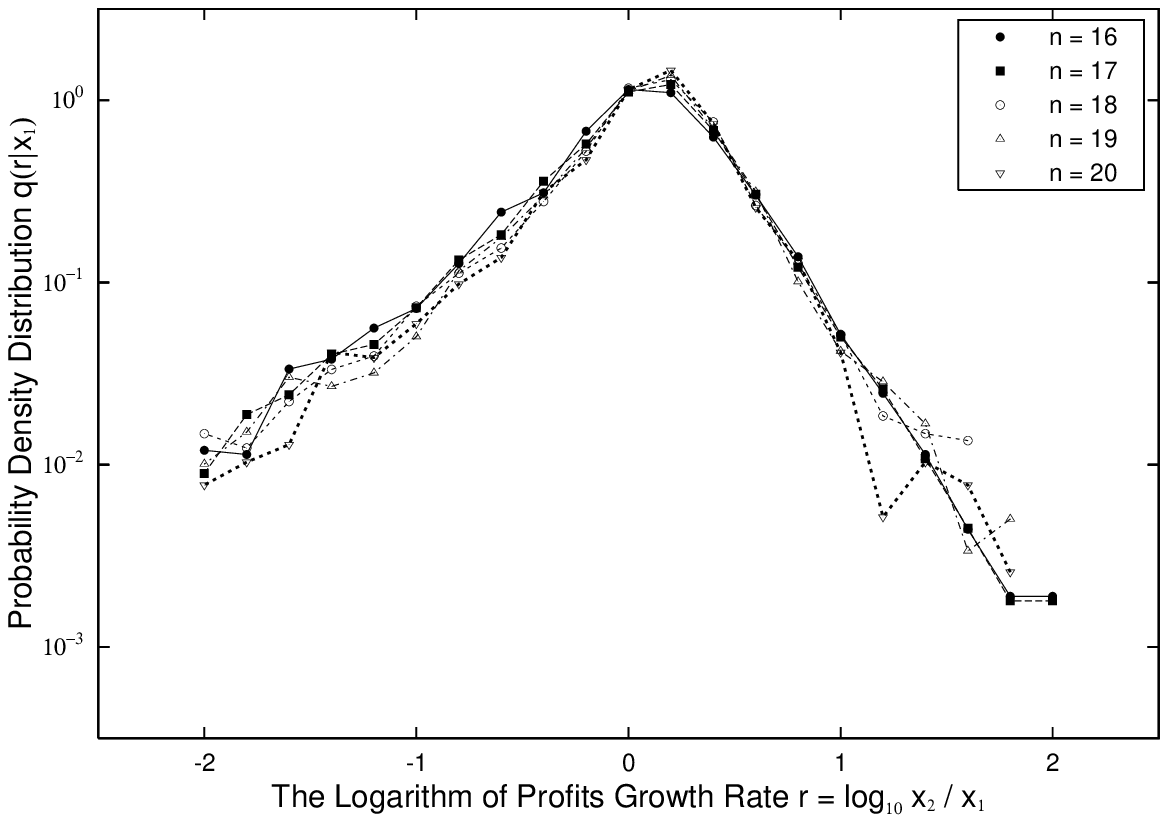}
 \end{minipage}
 \caption{The probability density distribution $q(r|x_1)$ of the log profits growth rate 
 $r = \log_{10} x_2/x_1$ 
 from $2003$ to $2005$ for $n=1, 2, \cdots, 20$.}
 \label{ProfitGrowthRate0305}
\end{figure}

\begin{figure}[htb]
%%%%%%%%%%%%%%%%%%%%%%%%%%%%%%%%%%%%%%%
%%%%%%%% 1st Line (two pictures)%%%%%%%
%%%%%%%%%%%%%%%%%%%%%%%%%%%%%%%%%%%%%%%
 \begin{minipage}[htb]{0.49\textwidth}
  \epsfxsize = 1.0\textwidth
  \epsfbox{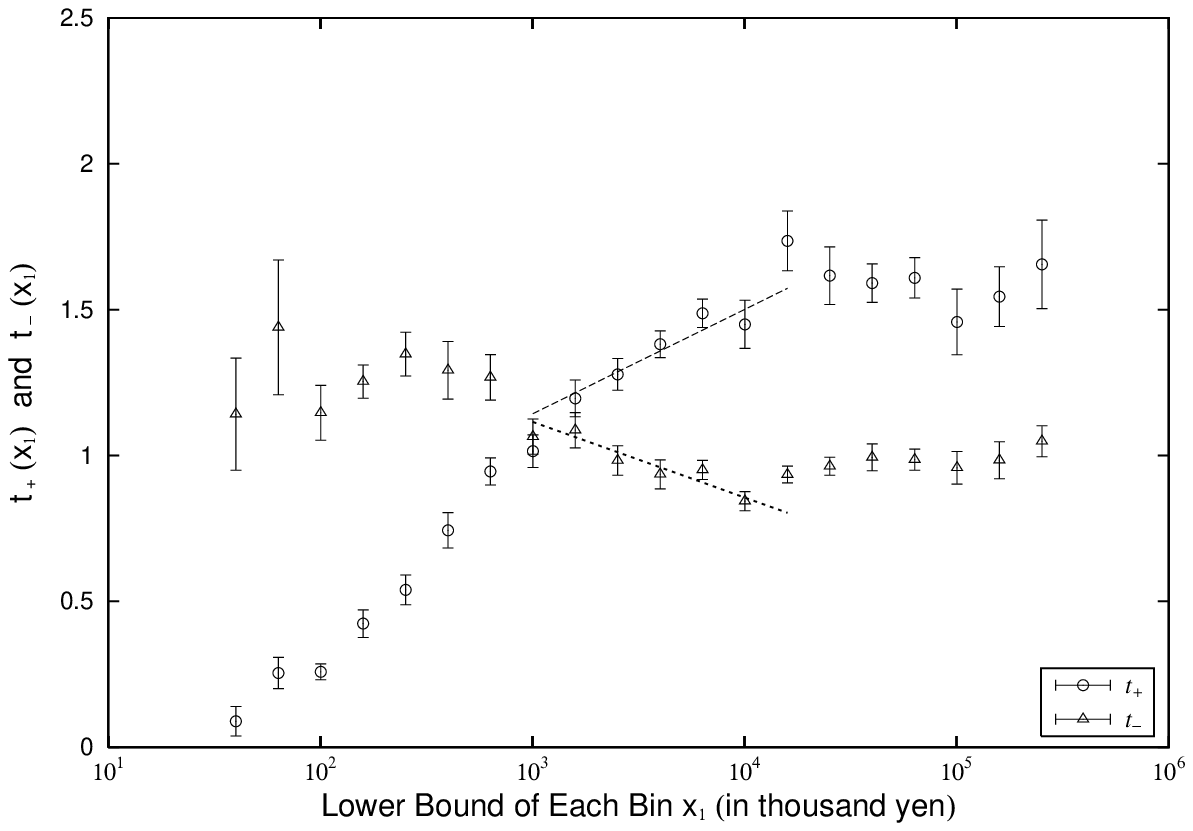}
 \caption{The relation between the lower bound of each bin $x_1$ and $t_{\pm}(x_1)$
 with respect to the profits growth rate from $2003$ to $2005$.
 From the left, each data point represents $n=1, 2, \cdots, 20$.}
 \label{eGibrat0305}
 \end{minipage}
 \hfill
 \begin{minipage}[htb]{0.49\textwidth}
  \epsfxsize = 1.0\textwidth
  \epsfbox{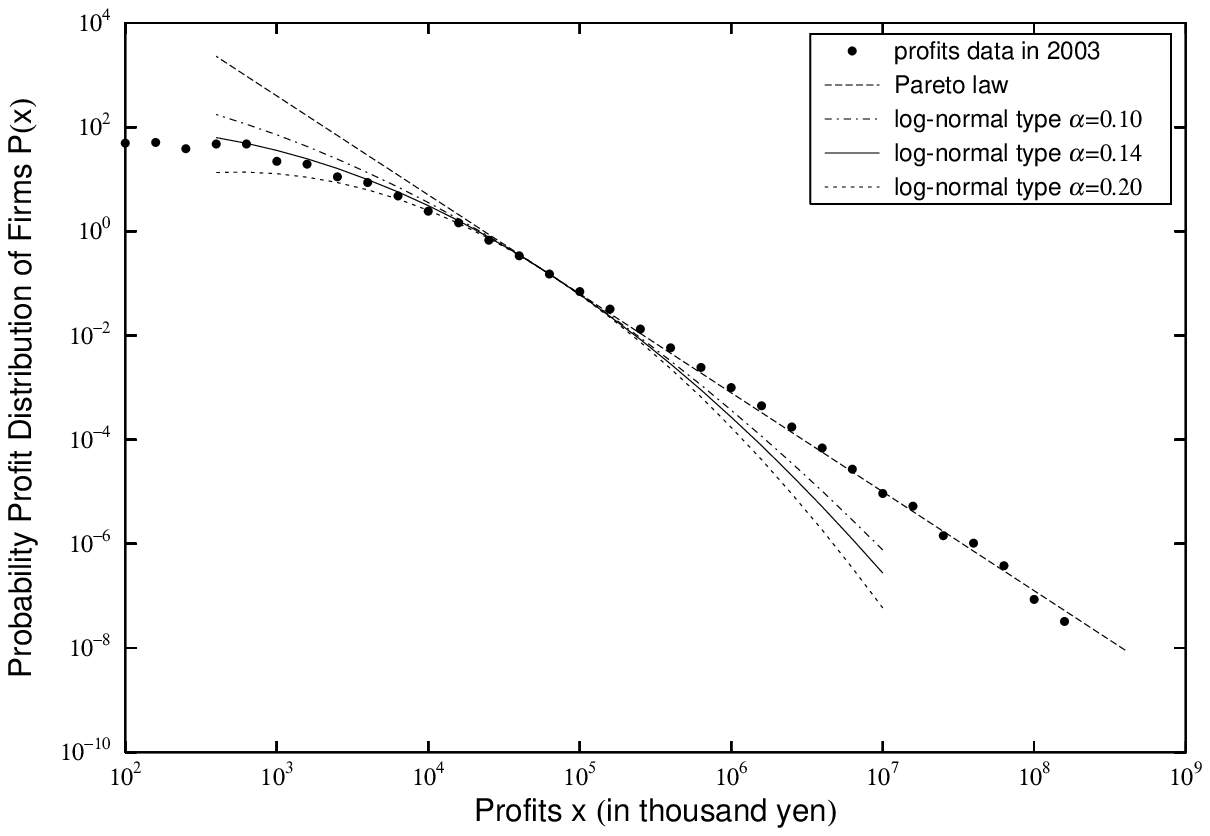}
 \caption{The probability distribution function of profits $P(x_1)$ for firms,
 the profits of which in 2003 ($x_1$) and 2005 ($x_2$) exceeded $0$,
 $x_1 >0$ and $x_2 > 0$.
 }
 \vspace{6.5mm}
 \label{Distribution0305-03}
 \end{minipage}
\end{figure}

These empirical data analyses are not restricted in the single term 2003--2004.
Figures~\ref{ProfitGrowthRate0405}--\ref{Distribution0405-04} show
similar empirical data analyses in 2004--2005.
The number of the firms in Fig.~\ref{ProfitGrowthRate0405}
is ``$20,669$", ``$89,064$", ``$88,651$" and ``$26,887$", respectively.
Furthermore, empirical data analyses in 2003--2005 are shown
in Figs.~\ref{ProfitGrowthRate0305}--\ref{Distribution0305-03}.
The number of the firms in Fig.~\ref{ProfitGrowthRate0305}
is ``$18,101$", ``$76,050$", ``$75,451$" and ``$22,456$", respectively.
In Figs.~\ref{eGibrat0405}, \ref{eGibrat0305},
the parameters $\alpha$, $x_0$ and $x_{\rm min}$ in 2004--2005 or 2003--2005
are estimated to be same values in 2003--2004.
The distribution (\ref{HandM}) in 2004--2005 or 2003--2005 also
fits with the empirical data consistently
(Figs.~\ref{Distribution0405-04}, \ref{Distribution0305-03}).

%%%%%%%%%%%%%%%%%%%%%%%%%%%%%%%%%%%%%%%%%%%%%%%%%
%%%       SECTION
%%%       Quasi-static log-normal distribution under detailed quasi-balance
%%%%%%%%%%%%%%%%%%%%%%%%%%%%%%%%%%%%%%%%%%%%%%%%%
\section{Quasi-static log-normal distribution under the detailed quasi-balance}
\label{Quasi-static log-normal distribution under the detailed quasi-balance}
\indent

In the previous section, 
the log-normal distribution in the middle scale region
is exhibited by the Non-Gibrat's law under the detailed balance.
The resultant profits distribution is empirically confirmed 
in data analyses of Japanese firms in 2003--2004, 2004--2005 and 2003--2005.
The profits distribution (\ref{HandM}) is static, 
because the derivation is based on the detailed balance (\ref{Detailed balance})
which is static time-reversal symmetry
(Fig.~\ref{logProfit0304DB}--\ref{logProfit0305DB}).
%In this section, we derive a Non-Gibrat's law under a detailed quasi-balance
%supposed in a quasi-static system~\cite{Ishikawa1}.

%In the derivation, we employ approximations confirmed in the static system. From 
%the Non-Gibrat's law under the detailed quasi-balance,
%we obtain a quasi-static log-normal distribution.

On the other hand,
we have derived Pareto's law with annually varying Pareto index
under the detailed quasi-balance~\cite{Ishikawa1}:
\begin{eqnarray}
    P_{1 2}(x_1, x_2) 
    = P_{1 2}( \left( \frac{x_2}{a} \right)^{1/{\theta}}, a~{x_1}^{\theta})~.
    \label{Detailed quasi-balance}
\end{eqnarray}
It is assumed that, in the quasi-static system,
the joint probability density function
has ``$a~{x_1}^{\theta} \leftrightarrow x_2$" symmetry
where $\theta$ is a slope of a regression line: 
\begin{eqnarray}
    \log_{10} x_2 = \theta~\log_{10} x_1 + \log_{10} a~.
    \label{Line}
\end{eqnarray}
The detailed balance (\ref{Detailed balance}) 
has the special symmetry $\theta = a = 1$.
Because the detailed quasi-balance (\ref{Detailed quasi-balance})
is imposed on the system,
$\theta$ is related to $a$ as follows:
\begin{eqnarray}
    \theta = 1 - \frac{2}{\Gamma} \log_{10} a~.
    \label{Gamma0}  
\end{eqnarray}
Here $10^{\Gamma}$ is a sufficient large value compared to the upper bound 
in which $\theta$ and $a$ are estimated.

In Ref.~\cite{Ishikawa1},
these results have been empirically confirmed 
by employing data on the assessed value of land~\cite{Kaizoji}--\cite{Web} 
in 1983--2005.
In the derivation,
we have used the Gibrat's law (\ref{Gibrat}) valid only in the large scale region
without linear approximations (\ref{approximation1})--(\ref{approximation2}).
The purpose of this section is to show that the 
approximations uniquely fix
a Non-Gibrat's law under the detailed quasi-balance.
After that, we identify the quasi-static distribution not only in the large scale region
but also in the middle scale region.

By using the relation $P_{1 2}(x_1, x_2)dx_1 dx_2 = P_{1 R}(x_1, R)dx_1 dR$
under the change of variables $(x_1, x_2)$ $\leftrightarrow  (x_1, R)$,
these two joint probability density functions are related to each other
\begin{eqnarray}
    P_{1 R}(x_1, R) = a~{x_1}^{\theta}~P_{1 2}(x_1, x_2)~,
\end{eqnarray}
where we use a modified growth rate $R= x_2/(a~{x_1}^{\theta})$~. From this relation,
the detailed quasi-balance (\ref{Detailed quasi-balance}) is rewritten as 
\begin{eqnarray}
    P_{1 R}(x_1, R) = R^{-1} P_{1 R}(\left(\frac{x_2}{a}\right)^{1/\theta}, R^{-1})~.
\end{eqnarray}
Substituting $P_{1 R}(x_1, R)$ for $Q(R|x_1)$ defined in Eq.~(\ref{define}),
the detailed quasi-balance is reduced to be
\begin{eqnarray}
    \frac{P(x_1)}{P(\left(x_2/a \right)^{1/\theta})} 
    = \frac{1}{R} \frac{Q(R^{-1}|\left(x_2/a \right)^{1/\theta})}{Q(R|x_1)}~.
    \label{}
\end{eqnarray}

In the quasi-static system, we also assume 
that $Q(R|x_1)$ follows the tent-shaped exponential forms 
(\ref{tent-shaped1})--(\ref{tent-shaped2}).
Under the approximations, the detailed quasi-balance is expressed as
\begin{eqnarray}
    \frac{\tilde{P}(x_1)}{\tilde{P}(\left(x_2/a \right)^{1/\theta})} 
    = R^{~t_{+}(x_1) - t_{-}(\left(x_2/a \right)^{1/\theta}) + 1}~
    \label{Detailed quasi-balance 2}
\end{eqnarray}
for $R>1$. Here we use the notation $\tilde{P}(x)=P(x) d(x)$.
By expanding Eq.(\ref{Detailed quasi-balance 2}) around $R=1$, 
the following differential equation is obtained:
\begin{eqnarray}
    \theta~\Bigl[t_{+}(x) - t_{-}(x) + 1 \Bigr] \tilde{P}(x) + x \tilde{P}~{'}(x) = 0~,
    \label{DE}
\end{eqnarray}
where $x$ denotes $x_1$.
The same equation is obtained for $R<1$.
The solution is expressed as 
\begin{eqnarray}
    \tilde{P}(x) = C x^{-\theta}~e^{-G(x)}~,
\label{solution}
\end{eqnarray}
where $\theta~\Bigl[t_{+}(x) - t_{-}(x) \Bigr] \equiv g(x)$ and
$\int dx~g(x)/x \equiv G(x)$~.

In order to make the solution (\ref{solution}) around $R=1$ satisfy 
Eq.~(\ref{Detailed quasi-balance 2}),
the following equation must be valid for all $R$:
\begin{eqnarray}
    - G(x) + G(R^{1/\theta} x) = \left[t_{+}(x) - t_{-}(R^{1/\theta} x) \right] \ln R~.
    \label{DE2}
\end{eqnarray}
The derivative of this equation with respect to $x$ is
\begin{eqnarray}
    - \frac{g(x)}{x} + \frac{g(R^{1/\theta} x)}{x} 
    = \left[{t_{+}}'(x) - R^{1/\theta}~{t_{-}}'(R^{1/\theta} x) \right] \ln R~.
    \label{derivative}
\end{eqnarray}
By expanding Eq.~(\ref{derivative}) around $R=1$, 
the following differential equations are obtained:
\begin{eqnarray}
    &&x~\Bigl[{t_{+}}^{''}(x)+{t_{-}}^{''}(x) \Bigr]
        +{t_{+}}^{'}(x)+{t_{-}}^{'}(x)=0~,\\
    &&(1-3\theta)~{t_{+}}^{'}(x)+(2-3\theta)~{t_{-}}^{'}(x)
    +3(1-\theta)~x~{t_{+}}^{''}(x)+3(2-\theta)~x~{t_{-}}^{''}(x)
    \nonumber\\
        &&~~~~~+x^2~\Bigl[{t_{+}}^{(3)}(x)+2~{t_{-}}^{(3)}(x) \Bigr]=0~.   
\end{eqnarray}
The solutions are given by
\begin{eqnarray}
    t_+(x) &=& -\frac{C_{-2}}{2} \ln^2 x 
        + \left(C_{+1}-C_{-1} \right) \ln x + \left( C_{+0}-C_{-0} \right)~,\\
    t_-(x) &=& \frac{C_{-2}}{2} \ln^2 x + C_{-1} \ln x + C_{-0}~.
\end{eqnarray}

In order to make these solutions satisfy Eq.~(\ref{DE2}),
the coefficients $C_{-2}$ and $C_{+1}$ must be $0$~.
As a result, $t_{\pm}(x)$ under the detailed quasi-balance is uniquely fixed as
\begin{eqnarray}
    t_{\pm}(x)=t_{\pm}(x_0) \pm \alpha~\ln \frac{x}{x_0}~.
    \label{t}
\end{eqnarray}
This is the same expression under the detailed balance.

By using the Non-Gibrat's law (\ref{t}),
probability density functions $P_1(x_1)$, $P_2(x_2)$ are uniquely reduced to be
\begin{eqnarray}
    P_1(x_1) &=& C_1 ~{x_1}^{-\mu_1-1}~
        \exp\left[-\theta~\alpha \ln^2 \frac{x_1}{x_0} \right],~~~
    \label{HandM1}\\
    P_2(x_2) &=& C_2 ~{x_2}^{-\mu_2-1}~
        \exp\left[-\theta~\alpha \ln^2 \frac{\left(x_2/a \right)^{1/\theta}}{x_0}\right]~
    \label{HandM2}
\end{eqnarray}
with
\begin{eqnarray}      
    \frac{\mu_1+1}{\mu_2+1} &=& \theta~
    \label{}  
\end{eqnarray}
in the large or middle scale
region where the dependence of $d(x_1)$ on $x_1$ is negligible.

Here we consider two log-normal distributions
in the middle scale region:
\begin{eqnarray}
    P_{\rm LN_1}(x_1) = \frac{1}{x_1 \sqrt{2 \pi {\sigma_1}^2}} \exp 
        \left[-\frac{\ln^2 \left( x_1/\bar{x}_1 \right)}{2 {\sigma_1}^2} \right]~,
    \label{LN1}\\
    P_{\rm LN_2}(x_2) = \frac{1}{x_2 \sqrt{2 \pi {\sigma_2}^2}} \exp 
        \left[-\frac{\ln^2 \left( x_2/\bar{x}_2 \right)}{2 {\sigma_2}^2} \right]~.
    \label{LN2}        
\end{eqnarray}
By comparing Eqs.~(\ref{HandM1})--(\ref{HandM2}) to (\ref{LN1})--(\ref{LN2}), we identify
\begin{eqnarray}
    \theta~\alpha = \frac{1}{2 {\sigma_1}^2}~,
    ~~~\mu_1 = \frac{1}{{\sigma_1}^2} \ln \frac{x_0}{\bar{x}_1}~,\\
    ~~~\frac{\alpha}{\theta} = \frac{1}{2 {\sigma_2}^2}~,
    ~~~\mu_2 = \frac{1}{{\sigma_2}^2} \ln \frac{a {x_0}^{\theta}}{\bar{x}_1}~.
\end{eqnarray}
Consequently, the relation between $\sigma_1$, $\sigma_2$ and $\theta$ is expressed as
\begin{eqnarray}
    \frac{\sigma_2}{\sigma_1} = \theta~.
\end{eqnarray}
This is the equation which quasi-static log-normal distributions satisfy.

%%%%%%%%%%%%%%%%%%%%%%%%%%%%%%%%%%%%%%%%%%%%%%
%%%       SECTION
%%%       Conclusion
%%%%%%%%%%%%%%%%%%%%%%%%%%%%%%%%%%%%%%%%%%%%%%
\section{Conclusion}
\label{sec-Conclusion}
\indent

In this study,
by employing profits data of Japanese firms,
we have exhibited the static log-normal distribution of profits in the middle scale region
from a Non-Gibrat's law under the detailed balance.
In the derivation, we have adopted two approximations
observed in the database.
One is that the probability density function of the profits growth rate is described as 
a tent-shaped exponential function.
The other is that the value of the origin of the growth rate distribution
divided into bins is constant. 
These approximations are confirmed in 2003--2004, 2004--2005 and 2003--2005.
The resultant profits distribution fits with the empirical data consistently.
This guarantees the validity of the approximations.

This static discussion has been applied to a quasi-static system.
We have derived a quasi-static log-normal distribution in the middle scale region
from a Non-Gibrat's law under the detailed quasi-balance.
In the derivation, we have 
also assumed two approximations confirmed in the database.
Even in the quasi-static system,
the approximations uniquely fix the Non-Gibrat's law to be the same expression
under the detailed balance.
The resultant distribution is power-law with varying Pareto index 
in the large scale region
and the quasi-static log-normal distribution in the middle scale region.
Notice that not only the change of Pareto index $\mu$ 
but also the change of the variance of the log-normal distribution $\sigma$
depends on a parameter of the detailed quasi-balance $\theta$.
In other words, $\mu$ and 
$\sigma$ are related to each other.
This phenomenon is suggested in Ref.~\cite{Souma}.

First issue in the future is to confirm this quasi-static derivation
by data analyses.
For this aim,
long-term economic data must be investigated in the middle scale region
where dynamical transitions are observed.
We believe that the analytic results in this study are confirmed in such a database, 
if two approximations assumed in this paper are applicable.

%%%%%%%%%%%%%%%%%%%%%%%%%%%%%%%%%%%%%%%%%%%%%%
%%%%%%%%%%%%%%%%%%%%%%%%%%%%%%%%%%%%%%%%%%%%%%

\section*{Acknowledgments}
\indent

%The author would like to thank
%H.~Itoh and TOKYO SHOKO RESEARCH, LTD.
%for kindly providing high-quality data.
The author is 
%also 
grateful to 
the Yukawa Institute for Theoretical 
Physics at Kyoto University,
where this work was initiated during the YITP-W-05-07 on
``Econophysics II -- Physics-based approach to Economic and
Social phenomena --'',
and especially to 
Professor~H. Aoyama for the critical question
about my previous work.
Thanks are also due to  Dr.~Y. Fujiwara
for a lot of useful discussions and comments.

%------------------ Bibliography -------------------------

%--------------------------------------------------------------

\newpage
%%%%%%%%%%%%%%%%%%%%%%%%%%%%%%%%%%%%%%%%%%%%
%%%%%%%%%%%%% FIGURE ( PLOTS )%%%%%%%%%%%%%
%%%%%%%%%%%%%%%%%%%%%%%%%%%%%%%%%%%%%%%%%%%%%%%%%%%%%%%%%
%%%%%%%%%%%%%%%%%%%%%%%%%%%%%%%%%%%%%%%%%%%%%%%%%%%%%%%%%
%%%%%%%%%%%% End 2nd Line %%%%%%%%%%%%%
%%%%%%%%%%%% End 2nd Line %%%%%%%%%%%%%
%%%%%%%%%%%% End 2nd Line %%%%%%%%%%%%%
%%%%%%%%%%%%%%%%%%%%%%%%%%%%%%%%%%%%%%%%%%%%%%%%%%%%%%%%%%
%\clearpage
%%%%%%%%%%%%%%%%%%%%%%%%%%%%%%%%%%%%%%%%%%%%
%%%%%%%%%%%%%%%%%%%%%%%%%%%%%%%%%%%%%%%%%%%%%%%%%%%%%%%%%%
%%%%%%%%%%%%%%%%%%%%%%%%%%%%%%%%%%%%%%%%%%%%%%%%%%%%%%%%%%
%%%%%%%%%%%%%%%%%%%%%%%%%%%%%%%%%%%%%%%%%%%%%%%%%%%%%%%%%

\begin{thebibliography}{99}
\bibitem{Gibrat}
    R.~Gibrat, Les inegalites economiques, Paris, Sirey, 1932.
\bibitem{Badger}
    W.W.~Badger, in: B.J.~West (Ed.), Mathematical Models as a Tool for the Social Science, 
    Gordon and Breach, New York, 1980, p. 87;\\
    E.W.~Montrll, M.F.~Shlesinger, J. Stat. Phys. 32 (1983) 209. 
\bibitem{Pareto}
    V.~Pareto, Cours d'Economique Politique, Macmillan, London, 1897.
\bibitem{Yakovenko}
    A.~Dr$\check{a}$gulescu, V.M.~Yakovenko, Physica A299 (2001) 213;\\
    A.C.~Silva, V.M.~Yakovenko, Europhys. Lett. 69 (2005) 304.
\bibitem{AIST}
    M.~Anazawa, A.~Ishikawa, T.~Suzuki and M.~Tomoyose, Physica A335 (2004) 616;\\
    A.~Ishikawa and T.~Suzuki, Physica A343 (2004) 376.   
\bibitem{Stanley1}
    M.H.R.~Stanley, L.A.N.~Amaral, S.V.~Buldyrev, S.~Havlin, H.~Leschhorn, 
    P.~Maass, M.A.~Salinger, H.E.~Stanley,
    Nature 379 (1996) 804.
%    L.A.N.~Amaral, S.V.~Buldyrev, S.~Havlin, H.~Leschhorn, P.~Maass,
%    M.A.~Salinger, H.E.~Stanley, M.H.R.~Stanley, J. Phys. (France) I7 (1997) 621;\\
%    S.V.~Buldyrev, L.A.N.~Amaral, S.~Havlin, H.~Leschhorn, P.~Maass,
%    M.A.~Salinger,  H.E.~Stanley, M.H.R.~Stanley, J. Phys. (France) I7 (1997) 635;\\
%    L.A.N.~Amaral, S.V.~Buldyrev, S.~Havlin, M.A.~Salinger, H.E.~Stanley, 
%    Phys. Rev. Lett. 80 (1998) 1385;\\
%    Y.~Lee, L.A.N.~Amaral, D.~Canning, M.~Meyer, H.E.~Stanley,
%    Phys. Rev. Lett. 81 (1998) 3275;\\
%    D.~Canning, L.A.N.~Amaral, Y.~Lee, M.~Meyer, H.E.~Stanley,
%    Economics Lett. 60 (1998) 335. 
\bibitem{TTOMS}
    H.~Takayasu, M.~Takayasu, M.P.~Okazaki, K.~Marumo, T.~Shimizu, cond-mat/0008057,
    in: M.M.~Novak (Ed.), Paradigms of Complexity, World Scientific, 2000, p. 243.
\bibitem{Aoyama}
    H.~Aoyama, 
    Ninth Annual Workshop on Economic Heterogeneous Interacting Agents (WEHIA 2004);\\
    H.~Aoyama, Y.~Fujiwara, W.~Souma,
    The Physical Society of Japan 2004 Autumn Meeting.
\bibitem{Ishikawa2007}
    A.~Ishikawa, 
    The uniqueness of firm size distribution function from tent-shaped growth rate distribution,
    physics/0702248.
\bibitem{FSAKA}
    Y.~Fujiwara, W.~Souma, H.~Aoyama, T.~Kaizoji and M.~Aoki,
    Physica A321 (2003) 598;\\
%    H.~Aoyama, W.~Souma and Y.~Fujiwara, Physica A324 (2003) 352;\\
    Y.~Fujiwara, C.D.~Guilmi, H.~Aoyama, M.~Gallegati and W.~Souma,
    Physica A335 (2004) 197.
%    Y.~Fujiwara, H.~Aoyama, C.D.~Guilmi, W.~Souma and M.~Gallegati,
%    Physica A344 (2004) 112;\\
%    H.~Aoyama, Y.~Fujiwara and W.~Souma, Physica A344 (2004) 117.
\bibitem{Ishikawa1}
    A.~Ishikawa, 
%    Annual change of Pareto index dynamically deduced from the law of detailed quasi-balance, 
    Physica A371 (2006) 525.
\bibitem{TSR}
    TOKYO SHOKO RESEARCH, LTD., http://www.tsr-net.co.jp/.
\bibitem{Ishikawa3}
    A.~Ishikawa, 
%    Derivation of the distribution from extended Gibrat's law,
    Physica A367 (2006) 425.
\bibitem{Ishikawa0}
    A.~Ishikawa, 
%    Pareto index induced from the scale of companies, 
    Physica A363 (2006) 367.
\bibitem{Kaizoji}
    T.~Kaizoji, Physica A326 (2003) 256;\\
    T.~Kaizoji and M.~Kaizoji, Physica A344 (2004) 138.
\bibitem{Web}
    The Ministry of Land, Infrastructure and Transport Government of Japan's World-Wide Web site,
    http://nlftp.mlit.go.jp/ksj/.
\bibitem{Souma}
    W.~Souma, Fractals 9 (2001) 463.
%\bibitem{ASNOTT}
%    H.~Aoyama, W.~Souma, Y.~Nagahara, H.P.~Okazaki, H.~Takayasu and M.~Takayasu, 
%    cond-mat/0006038, Fractals 8 (2000) 293;\\
%    W.~Souma, cond-mat/0011373, Fractals 9 (2001) 463.
%\bibitem{OTT}
%    K.~Okuyama, M.~Takayasu and H.~Takayasu,
%    Physica A269 (1999) 125.
%\bibitem{Mizuno}
%    T.~Mizuno, M.~Katori, H.~Takayasu and M.~Takayasu, cond-mat/0308365,
%    in: H.~Takayasu (Ed.), Empirical Science of Financial Fluctuations: The Advent of Econophysics, 
%    vol. 321, Springer, Tokyo, 2003.
%\bibitem{ASF}
%    H.~Aoyama, W.~Souma and Y.~Fujiwara, Physica A324 (2003) 352.
%\bibitem{Ishikawa}
%    A.~Ishikawa, cond-mat/0409145, Physica A349 (2005) 597.
%\bibitem{Zipf}
%    G.K.~Gipf, Human Behavior and the Principle of Least Effort, Addison-Wesley, Cambridge, 1949.
%\bibitem{MS}
%    R.N.~Mategna and H.E.~Stanley, An Introduction to Econophysics, 
%    Cambridge University Press, UK, 2000.
%\bibitem{Stanley1}
%    M.H.R.~Stanley, L.A.N.~Amaral, S.V.~Buldyrev, S.~Havlin, H.~Leschhorn, P.~Maass, M.A.~Salinger and H.E.~Stanley,

%    Nature 379 (1996) 804;\\
%    L.A.N.~Amaral, S.V.~Buldyrev, S.~Havlin, H.~Leschhorn, P.~Maass,
%    M.A.~Salinger, H.E.~Stanley and M.H.R.~Stanley, J. Phys. (France) I7 (1997) 621;\\
%    S.V.~Buldyrev, L.A.N.~Amaral, S.~Havlin, H.~Leschhorn, P.~Maass,
%    M.A.~Salinger,  H.E.~Stanley and M.H.R.~Stanley, J. Phys. (France) I7 (1997) 635;\\
%    L.A.N.~Amaral, S.V.~Buldyrev, S.~Havlin, M.A.~Salinger and H.E.~Stanley, 
%    Phys. Rev. Lett. 80 (1998) 1385;\\
%    Y.~Lee, L.A.N.~Amaral, D.~Canning, M.~Meyer and H.E.~Stanley,
%    Phys. Rev. Lett. 81 (1998) 3275;\\
%    D.~Canning, L.A.N.~Amaral, Y.~Lee, M.~Meyer and H.E.~Stanley,
%    Economics Lett. 60 (1998) 335.
\end{thebibliography}
\end{document}